\documentclass[galaxies,review,submit,moreauthors,pdftex]{Definitions/mdpi} 
    \usepackage{graphicx,color}
    \usepackage{amsmath}
    \usepackage{natbib}
    \usepackage{amssymb}
    \usepackage{bm}
    \usepackage{ulem}
    \usepackage{upgreek}
    \usepackage{enumitem}
    \usepackage{subcaption}
    
    \setlist{nosep}	

    \newcommand{\Del}{{\nabla}}
    \newcommand{\bfDel}{\bm{\nabla}}

    \newcommand{\bfU}{\bm{U}}
    \newcommand{\bfB}{\bm{B}}
    \newcommand{\bfu}{\bm{u}}
    
    \newcommand{\bfb}{\bm{b}}

    \newcommand{\mean}[1]{\overline{#1}}
    \newcommand{\meanv}[1]{\overline{\bm{#1}}}
    
    \newcommand{\eq}{_\mathrm{eq}}						
    \newcommand{\f}{_\mathrm{0}}					   	
    \newcommand{\magn}{_\mathrm{m}}			   		
    \newcommand{\turb}{_\mathrm{t}}			   		
    \newcommand{\etat}{\eta_\mathrm{t}}			   	
    \newcommand{\crit}{_\mathrm{c}}			   		
    
    \newcommand{\cro}{\times}

    \newcommand{\mbr}{\mean{B}_r}
    \newcommand{\mbp}{\mean{B}_\phi}
    \newcommand{\mbz}{\mean{B}_z}

    \newcommand{\pat}{_\mathrm{p}}

    \newcommand{\disk}{_\mathrm{d}}
    
    \newcommand{\halo}{_\mathrm{h}}

    \newcommand{\shear}{_\mathrm{s}}
    \newcommand{\eddy}{_\mathrm{e}}
    
    \def\hi{H\,{\sc i}}
    \def\msun{$M_{\odot}$}

    \makeatletter
    \newsavebox{\@brx}
    
    \newcommand{\llangle}[1][]{\savebox{\@brx}{\(\m@th{#1\langle}\)}%
      \mathopen{\copy\@brx\kern-0.5\wd\@brx\usebox{\@brx}}}
    \newcommand{\rrangle}[1][]{\savebox{\@brx}{\(\m@th{#1\rangle}\)}%
      \mathclose{\copy\@brx\kern-0.5\wd\@brx\usebox{\@brx}}}
    
    \newcommand{\kms}{\,{\rm km\,s^{-1}}}
    \newcommand{\kmskpc}{\,{\rm km\,s^{-1}\,kpc^{-1}}}

    \newcommand{\kpc}{\,{\rm kpc}}
    \newcommand{\pc}{\,{\rm pc}}
    
    \newcommand{\Myr}{\,{\rm Myr}}
    
    \newcommand{\Gyr}{\,{\rm Gyr}}

    \newcommand{\mkG}{\,\mu{\rm G}}

    \newcommand{\yr}{\,{\rm yr}}

    \firstpage{1} 
    \pubvolume{xx}
    \issuenum{1}
    \articlenumber{5}
    \pubyear{2019}
    \copyrightyear{2019}
    \history{Received: 10 October 2019; Accepted: 16 December 2019; Published: Jan 2020}
    
    \pdfoutput=1
    
    \Title{Synthesizing Observations and Theory to Understand Galactic Magnetic Fields: Progress and Challenges}
    
    \Author{Rainer Beck $^{1}$*\orcidA{}, Luke Chamandy $^{2}$\orcidB{}, Ed Elson $^{3}$\orcidC{}, and Eric G. Blackman $^{2}$\orcidD{}}
    
    \AuthorNames{Rainer Beck, Luke Chamandy, Ed Elson, and Eric G. Blackman}
    
    \address{$^{1}$ \quad Max-Planck-Institut f\"ur Radioastronomie, Auf dem H\"ugel 69, 53121 Bonn, Germany\\
    $^{2}$ \quad Department of Physics and Astronomy, University of Rochester, Rochester NY 14627, USA\\
    $^{3}$ \quad Department of Physics, University of the Western Cape, Belleville 7535, Republic of South Africa}
    
    \corres{Correspondence: rbeck@mpifr-bonn.mpg.de}

    \abstract{
    Constraining dynamo theories of magnetic field origin  by observation is indispensable but challenging,
    in part because the basic quantities measured by observers and predicted by modelers are different. 
    We clarify these differences and sketch out ways to bridge the divide.
    Based on archival and previously unpublished data, we then compile various important properties of galactic magnetic fields for nearby spiral galaxies. We consistently compute 
    strengths of total, ordered, and regular fields, pitch angles of ordered and regular fields, and we summarize the present knowledge on azimuthal modes, field parities, and the properties of non-axisymmetric spiral features called magnetic arms.
    We review related aspects of dynamo theory, 
    with a focus on mean-field models and their predictions for large-scale magnetic fields in galactic discs and halos.
    Further, we measure the velocity dispersion of \hi\ gas in arm and inter-arm regions in three galaxies,
    M\,51, M\,74, and NGC\,6946, since spiral modulation of the root-mean-square turbulent speed
    has been proposed as a driver of non-axisymmetry in large-scale dynamos.
    We find no evidence for such a modulation and place upper limits on its strength,
    helping to narrow down the list of mechanisms to explain magnetic arms.
    Successes and remaining challenges of dynamo models with respect to explaining observations 
    are briefly summarized, and possible strategies are suggested.
    With new instruments like the Square Kilometre Array (SKA),
    large data sets of magnetic and non-magnetic properties from thousands of galaxies will become available, 
    to be compared with theory.
    }
    
    
\begin{document}
    
    
    \section{Introduction}
    \label{sec:intro}
    
    The presence of magnetic fields in the interstellar medium (ISM) of the Milky Way and  external galaxies has now been known for more than 60 years. Their presence raises four natural questions:  (1) How did they get there? (2) What is their structure? 
    (3) Are they dynamically influential? (4) What might we learn from their properties about galaxy structure or dynamical processes?
    Fully answering these questions covers very broad ground and although we touch on aspects of all of them, we focus here on the the curious fact that the magnetic field of spiral galaxies often exhibits a large-scale component with net flux of one sign over a large portion of the galactic area, even when there is a smaller scale random component.
    How does such a large-scale field arise and survive in galaxies amidst the otherwise turbulent and chaotic interstellar media?
    
    Since large-scale magnetic fields are subject to turbulent diffusion  and the vertical diffusion time scale is typically
    10 times less than the age of the Universe \citep[e.g.][]{Chamandy+14b}, these fields must be replenished in situ, regardless of whatever seed field (primordial or protogalactic) may have been supplied. Since the diffusion represents exponential decay, the growth and replenishment must itself supply exponential growth. This fact has led to a grand enterprise of in-situ galactic dynamo theory and modeling to explain the large-scale fields. The purpose of our review is to bring the reader up to date on the efficacy with which large-scale dynamo theory and observation are consistent and where challenges remain.
    
    \subsection{Radio observations}
    \label{sec:radio_observations}
    
    Synchrotron emission from star-forming galaxies and hence evidence for magnetic fields 
    was first detected by \citet{Brown+Hazard51} in the nearby spiral galaxy M\,31. 
    Measurement of linearly polarized emission and its Faraday rotation needs sensitive receiving systems 
    and good telescope resolution and was first successfully achieved in the Milky Way in 1962 \citep{Westerhout+62,Wielebinski+62}, 
    in 1972 for the spiral galaxy M\,51 by \citet{Mathewson+72}, and in 1978 for M\,31 by \citet{Beck+78}. 
    The large-scale regular field in M\,31 discovered by \citet{Beck82} 
    gave the first strong evidence for the action of a large-scale dynamo. 
    Since then, several hundred
    star-forming galaxies have been mapped with various radio telescopes. 
    The combination of total intensity and polarization data from high-resolution interferometric (synthesis) telescopes 
    and data from single-dish telescopes providing large-scale diffuse emission was particularly successful \citep[e.g.][]{Fletcher+11}. 
    A new method, called rotation measure synthesis, based on the seminal work by \citet{Burn66}, was introduced in 2005 by \citet{Brentjens+05} 
    and allows measurement of Faraday rotation and intrinsic polarization angles with a single broadband receiver.
    
    Total synchrotron emission indicates the presence of magnetic fields in galactic discs and halos 
    and allows us to estimate the total field strength, 
    if the conventional 
    assumption of energy density equipartition between magnetic fields and cosmic rays is valid (see Section~\ref{sec:strength_B_tot}).
    Comparisons with large-scale dynamo theory are based on linearly
    polarized emission and its Faraday rotation.
    Linear polarization has been found in more than 100 galaxies \citep[][and updates on arXiv]{Beck+Wielebinski13}, 
    while the detection of Faraday rotation needs multi-frequency observations and high spatial resolution. 
    Systematic investigations have been performed for about 20 galaxies for which sufficiently detailed data were available \citep{Vaneck+15} (see Section~\ref{sec:correlations}).
    Reviews of observational results were given by \citet{Fletcher10} and \citet{Beck16}.
        
    \subsection{Galactic dynamo theory and simulations}
    \label{sec:galactic_dynamo_theory}
    
    Mean-field or large-scale dynamo models are based on mean-field electrodynamics,
    wherein the magnetic and velocity fields are formally separated into vectors of mean and fluctuating parts, 
    i.e. $\bfB= \meanv{B} +\bfb$ and $\bfU= \meanv{U} +\bfu$, 
    with the mean of the fluctuations equal to zero 
    (mean quantities are denoted with bar and fluctuating quantities with lower case).
    We further discuss averaging and the connection between mean and large-scale in Sec.~\ref{sec:def},
    but ``large-scale'' and ``small-scale'' generally refer to scales larger and smaller 
    than the correlation length of turbulence.
    The theory requires solving for the mean field $\meanv{B}$ in terms of the mean 
    velocity field $\meanv{U}$ (usually prescribed) and correlations of the  small-scale turbulent fluctuations $\bfu$ and $\bfb$.
    The latter comprise turbulent transport tensors like $\boldmath\alpha$ and $\boldmath\etat$. 
    These quantities 
    can be estimated using theory or direct numerical simulations (DNSs) but can also be heuristic
    \citep{Moffatt78,Krause+Radler80,Brandenburg+Subramanian05a}.
    
    Early models focused on the linear or kinematic regime, 
    whose solutions are exponentially growing (or decaying) eigenmodes of the averaged induction equation.
    The non-linear regime begins as the energy density of the mean field  becomes comparable to the mean energy density of the turbulence,
    i.e. when the field strength $\mean{B}$ approaches the equipartition value
    \begin{equation}
      \label{Beq}
      B\eq= (4\pi\rho)^{1/2} u,
    \end{equation}
    where $\rho$ is the gas density and $u=(\mean{\bfu^2})^{1/2}$.
   This type of energy equipartition is separate from the energy equipartition between magnetic field and CRs 
    discussed in Sec.~\ref{sec:strength}.
    At this point, the backreaction of the Lorentz force on the flow becomes significant, and is expected to  quench the mean-field dynamo.
    
    Research has  focused on so-called $\alpha\Omega$ dynamos,
    for which differential rotation shears the mean poloidal field to produce a mean toroidal field (the ``$\Omega$ effect''),
    while the ``$\alpha$ effect'' transforms toroidal to poloidal fields, completing the feedback loop needed for exponential growth.
    The $\alpha$ effect is commonly taken proportional to the mean kinetic helicity density in the turbulent flow, 
    presumably resulting from the action of the Coriolis force on vertically moving fluid elements in the stratified ISM
    \citep{Parker55,Krause+Radler80,Ruzmaikin+88,Brandenburg+13}.
    The $\alpha$ effect can also help to make toroidal out of poloidal fields in a  more general so-called $\alpha^2\Omega$ dynamo, 
    but in galactic discs, where the $\Omega$ effect is strong, the $\alpha^2$ effect is likely subdominant.
    The $\alpha^2$ effect may be important in the central regions of galaxies, 
    where the global shear parameter $q\shear=-d\ln\Omega/d\ln r$ is small compared to $1$ \citep[e.g][]{Chamandy16}.
    
    An alternative view is that the $\alpha$ effect is supplied by the Parker instability rather than cyclonic turbulence. 
    This instability presumably originates in the disc 
    and leads to buoyant magnetic field loops that are twisted by the Coriolis force as they expand into the halo,
    converting toroidal field to poloidal field.
    CRs likely supply part of the buoyancy of these rising Parker loops. 
    Furthermore, the interaction of adjacent Parker loops may contribute to the field dissipation 
    \citep{Parker92,Hanasz+Lesch93,Hanasz+Lesch98}.
    \citet{Moss+99} presented a detailed mean-field galactic dynamo model based on these ideas. \citet{Kulsrud2015} discussed how problems with \cite{Parker92} in the early weak-field phase of the dynamo might be circumvented.
    Numerical simulations of the so-called ``cosmic-ray driven dynamo'' model were presented in \citet{Hanasz+04} and \citet{Hanasz+09b}.
    A recent study exploring the evolution of the Parker instability in galactic discs using DNSs was carried out by \citet{Rodrigues+16}. 
    
    A leading model for mean-field dynamo non-linearity is predicated on the principle of magnetic helicity conservation,
    and is known as dynamical $\alpha$-quenching
    \citep[][]{Pouquet+76,Kleeorin+Ruzmaikin82,Blackman+Field02,Blackman+Brandenburg02,Brandenburg+Subramanian05a,Subramanian+Brandenburg06}.
    Large-scale magnetic helicity generated by the mean-field dynamo must be compensated by oppositely signed small-scale helicity
    to conserve the total magnetic helicity.
    This small-scale helicity (more precisely, the related quantity, 
    current helicity) contributes a term $\alpha\magn$ to the $\alpha$ effect
    that has opposite sign to the kinetic $\alpha$ term, leading to a suppression of the total $\alpha$.
    To avoid ``catastrophic quenching'', there must be a flux of 
    $\alpha\magn$ outward from the dynamo-active region
    \citep{Kleeorin+Rogachevskii99,Blackman+Field00,Vishniac+Cho01,Blackman+Field02,Field+Blackman02}. 
    Dynamical $\alpha$-quenching of the mean field growth can be approximated by an older heuristic approach known as algebraic quenching \citep{Chamandy+14b}, 
    but the physical origin of $\alpha$-quenching comes from the  dynamical connection to magnetic helicity evolution.
    More details about galactic mean-field dynamo models can be found in Sec.~\ref{sec:strength_theory}.
    
    While asymptotic analytical solutions are possible, 
    the mean-field equations are usually solved numerically as an initial value problem. 
    In this case, models are sometimes referred to as mean-field simulations.
    These differ from DNSs, which solve the full MHD equations.
    Likewise, models may be local (limited to a small part of the ISM), global (modeling an entire galaxy), or cosmological.
    No study has yet come close to including the full dynamical range of scales of the galactic dynamo problem,
    and so all of these approaches are valuable.
    Previous reviews of galactic dynamos include Refs.~\citep{Ruzmaikin+88,Beck+96,Shukurov05,Shukurov07,Kulsrud+Zweibel08,schaye_2004,Brandenburg14}
    as well as Refs.~\citep{Moss+Sokoloff19,Subramanian19}.
    
    \subsection{Outline}
    
    The main goal of this work is to review the current status of magnetic field observations and dynamo models, 
    focusing on large-scale magnetic fields in the discs of nearby spiral galaxies. 
    Throughout, we present updated compilations of magnetic field data for the best-studied nearby galaxies.
    A highlight of our work, in Sec.~\ref{sec:veloc_disp_contrast}, 
    is a stand-alone effort to constrain directly dynamo parameters 
    by measuring the root-mean-square (rms) turbulent velocity in arm and inter-arm regions for three galaxies:
    we see this as an example of the sort of inter-disciplinary study that is needed to advance the field.
    
    We begin by laying out the definitions of the various magnetic field components in observations
    and in dynamo theory in Sec.~\ref{sec:def}, where we also highlight the challenges involved
    in getting observers and theorists to speak a common language.
    Sec.~\ref{sec:geometry} explores the geometry of the mean magnetic field,
    including sign, parity, and reversals.
    In Sec.~\ref{sec:strength} we discuss the strength of the magnetic field.
    The roles of the small-scale fluctuating field component and seed fields in mean-field dynamo models 
    are discussed in Sec.~\ref{sec:seed_ss}.
    Sec.~\ref{sec:pitch} is concerned with the pitch angles of the magnetic field in both observation and theory.
    We then briefly touch on statistical correlations between field properties in Sec.~\ref{sec:correlations}
    and on halo magnetic fields in Sec.~\ref{sec:halo}.
    In Sec.~\ref{sec:non-axisymmetric} we provide a fairly detailed review of non-axisymmetric magnetic fields.
    We conclude and present our outlook in Sec.~\ref{sec:conclusion}.
    We have chosen  not to cover in detail magnetic fields in the Milky Way, galaxies with strong bars,
    high-redshift galaxies, interacting galaxies, and dwarf galaxies.
    
    \section{Definitions of Magnetic Field Components}
    \label{sec:def}
    
    \subsection{Observations}
    \label{sec:obs_def}
    
\begin{figure*}[t]
     \vspace*{7mm}
     \begin{center}
     \includegraphics[width=10cm]{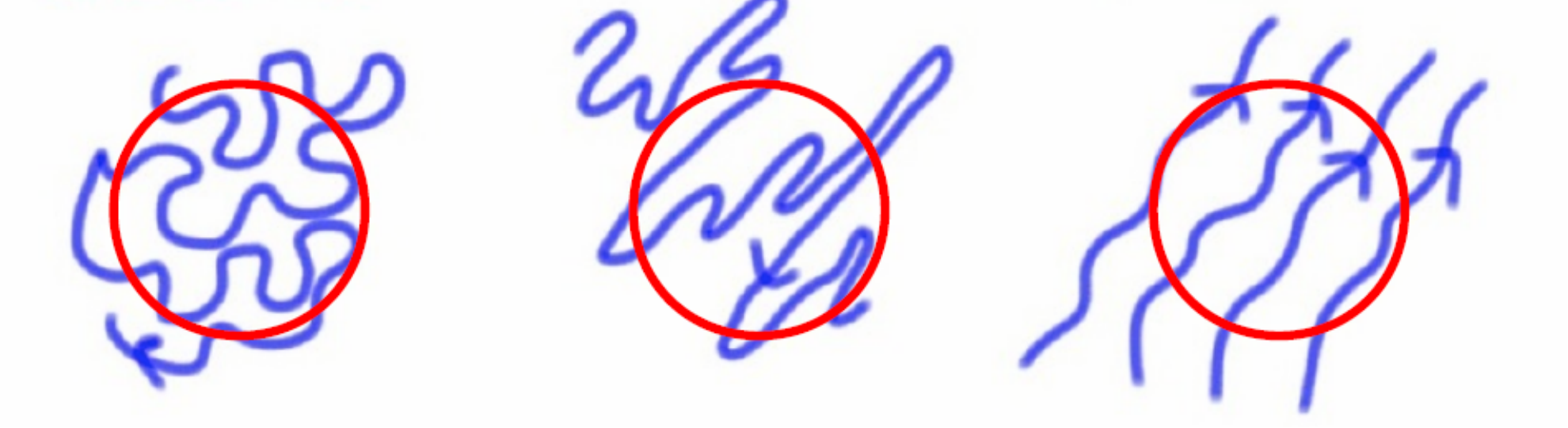}
     \caption{Schematic illustration of magnetic field components:
     mostly isotropic turbulent field ${\bfB_\mathrm{iso}}$ (left), mostly anisotropic turbulent field $\bfB_\mathrm{an}$ (middle), 
     and mostly regular field $\bfB_\mathrm{reg}$ (right).
     The red circle represents the telescope beam (courtesy: Andrew Fletcher).
     }
     \label{fig:cartoon}
     \end{center}
     \end{figure*}

    Observers separate the total magnetic field into a \textit{regular} and a \textit{turbulent} component. 
    A regular (or coherent) field has a well-defined direction within the beam width of the telescope, 
    while a turbulent field has one or more spatial reversals within the beam.
    Turbulent fields can be isotropic (i.e. the same dispersion in all three spatial dimensions) or anisotropic (i.e. different dispersions).
    Figure~\ref{fig:cartoon} shows three examples of field configurations; from left to right, the dominant contribution is from isotropic turbulent field, anisotropic turbulent field, or regular field.
    For more explanation of these different field components, along with a useful schematic diagram,
    see \citet{Jaffe19}, but note the different choice of nomenclature.
 
    Total synchrotron intensity traces the total magnetic field in the plane of the sky. 
    The ``ordered field''  $\bfB_\mathrm{ord}$ is defined to be what polarized synchrotron intensity 
    at high radio frequencies (to avoid Faraday depolarization) measures within the telescope beam, 
    projected to the plane perpendicular to the line of sight.
    \footnote{Other definitions of ``ordered field'' are used in the literature, e.g. in \cite{Jaffe19}.}
    The ``anisotropic turbulent field'' $\bfB_\mathrm{an}$, whose average over the beam vanishes, 
    and the ``regular field'' $\bfB_\mathrm{reg}$, whose average over the beam is finite,  both contribute to the ordered field.
    A large telescope beam  may not resolve small-scale field structure, 
    so  the observed radio emission will appear less  polarized than for a smaller beam.
    Unpolarized synchrotron intensity from external galaxies is attributed to an isotropic turbulent field ${\bfB_\mathrm{iso}}$.
    Observations cannot resolve fields with small-scale structure below the angular beam width of typically between $15''$ and $4'$, which corresponds to a few hundred\,pc in nearby galaxies.
    
    The polarization angle (corrected for Faraday rotation) shows the field \textit{orientation} in the plane of the sky, but
    with $n\times 180^\circ$ ambiguity, and hence is insensitive to field reversals 
    that occur on scales smaller than the telescope beam. 
    Faraday rotation (and the longitudinal Zeeman effect) is sensitive to the \textit{direction} of the field 
    along the line of sight and hence can unambiguously trace regular fields. 

    Depending on the task of investigation, averages of total and/or polarized synchrotron intensity are computed, 
    e.g. globally for the whole galaxy,  2$\pi$ azimuthal averages in annuli of a few kpc width, 
    or averages over sectors of a given radial and azimuthal range. 
    These averages are then transformed into average field strengths (Section~\ref{sec:strength}), 
    degree of polarization, polarization angle, and Faraday rotation measure.
    
    It is tempting to equate the regular field $\bfB_\mathrm{reg}$ with the mean field $\meanv{B}$ of dynamo theory, 
    and the combined vector $\bfB_\mathrm{an}+\bfB_\mathrm{iso}$ with the fluctuating field $\bfb$.
    However, the extent to which they are directly mutually transferable 
    is determined by the extent to which the implicit averaging of the mean-field model
    approximates the averaging in the observations for the system under consideration.


    \subsection{Theory}
    \label{sec:theory_def}
    
    The averaging procedure for deriving the standard mean-field equations employs certain mathematical rules 
    (the Reynolds averaging rules). Theorists typically think of the mean field $\meanv{B}$ as an ensemble average of $\bfB$
    over an infinite set of statistical realizations (or equivalently, averaging over infinite time for a field in the steady state) \citep[][Sec.VII.2]{Ruzmaikin+88}.
    Understood this way, the fluctuating field component $\bfb$ that is generated from random processes averages to zero,
    and $\bfB=\meanv{B}+\bfb$.
    Below we refer to this purely theoretical statistical ensemble averaged mean denoted by bar as the ``ensemble mean''.
    The ensemble mean has infinite precision in both space and time.
    
    To make contact with observations and DNSs, mean-field models can be constructed using spatial, 
    rather than ensemble averages \citep{Zhou+18}.
    A spatial mean approaches the ensemble mean only in the limit that (i) the turbulent correlation length
    is infinitely smaller than the scale of averaging \textit{and} (ii) the ensemble mean is uniform within the averaging kernel.
    Galaxies do not have turbulence correlation lengths $l$ infinitely smaller than the averaging scale $\mathcal{L}$.
    This mismatch leads to ``noise'' in spatially averaged quantities, typically of order $(l/\mathcal{L})^{3/2}$, 
    corresponding to the average contribution to the mean from $(\mathcal{L}/l)^3$ turbulent cells with randomly directed field.
    Taking into account solenoidality of the field reduces the estimate of the noise amplitude \citep[][VII.14]{Ruzmaikin+88}.
    In any case, this noise should be re-incorporated into the solution or model, leading to precision error in the theory.
    In addition, the variation scale of the ensemble mean 
    $|\meanv{B}|/|\nabla \meanv{B}|$
    is never infinitely larger than the averaging scale $\mathcal{L}$.
    This mismatch leads to systematic differences which depend on the ratio of these scales,
    and result in higher order terms in the mean-field equations.
    
    In general, comparing a mean-field model constructed using a given definition of the mean
    with observational data that stem from an implicit averaging filter (e.g. line of sight averaging to obtain the $RM$),
    requires that the model also be subjected to the filter.
    This second averaging leads to additional errors related to deviations from conditions (i) and (ii) \citep{Zhou+18}.
    
    DNSs solve the full (unaveraged) MHD equations.
    However, to make contact between DNSs and mean-field models or observations,
    choices must be made as to how to determine mean and random components a posteriori. 
    Possible choices include spatial averages, temporal averages, and spectral filters
    (see \citep{Shapovalov+Vishniac11} for a useful discussion on the latter).
    Care must be taken to make these choices realistic. 
    For example, planar and box averages for periodic boxes are unphysical \citep{Zhou+18}.
    Gaussian filtering provides a rigorous practical averaging procedure for making contact with observations \citep[e.g.][]{Gent+13b}.
    Because it allows for finite scale separation, it does not obey the Reynolds averaging rules of mean-field theory \citep{Germano92} 
    but facilitates quantifying this deviation \citep{Zhou+18} to quantify the distinction 
    between what observers measure vs. what theorists calculate.

    \section{Geometry of the Large-Scale Field}
    \label{sec:geometry}
    
    \subsection{Sign of the Field}
    \label{sec:sign}
    
    According to \citet{Krause+Beck98}, a comparison of the sign of Faraday rotation measures ($RM$) with that of the rotational velocity along the line of sight measured near the major axis of a galaxy allows determining the sign of radial component of the axisymmetric spiral field,
    under two conditions: (1) the large-scale field of a galaxy can be described by an axisymmetric mode along azimuthal angle in the galaxy plane ($m=0$) or by a combination of $m=0$ and higher modes, and (2) the spiral arms are trailing. 
    This method has been applied to 16 external galaxies so far (Table~\ref{tab:directions}).
    Figure~\ref{fig:m83} gives an example obtained from so far unpublished data of the barred spiral galaxy M\,83.
    The disc fields reveal similar occurrence of both radial signs (ratio 6:9). In one galaxy (NGC\,4666) and in the Milky Way the field direction changes within the disc.
    
    Three galaxies reveal a field reversal from outward in the central region to inward in the disc. As rotational velocities and/or inclinations of the central regions in these galaxies are different from those of the disc, they are probably dynamically decoupled.
    No case with the opposite sense of reversal has been found so far.
    The dynamo equations are invariant under a change of sign of the magnetic field,
    so no preference of one direction over the opposite direction is expected.
    Random processes determine which direction manifests.

    \begin{table*}
    \begin{center}
    \caption{Direction of the radial component of spiral fields in the discs of 15 spiral galaxies and one irregular dwarf galaxy (NGC\,4449) with a dominating axisymmetric mode.
    }
    \label{tab:directions}
    \vspace{0.2cm}
    \begin{tabular}{@{}lcc@{}}
    \hline
    Direction & Galaxies & References \\
    \hline
    Inward (disc) & M\,33, NGC\,253, NGC\,6946 & \cite{Tabatabaei+08,Heesen+09b,Beck07}\\
    Outward (central region) \& inward (disc)   & M\,31, NGC\,2997, IC\,342 & \cite{Berkhuijsen+97,Giessuebel+Beck14,Han+99,Beck15a}\\
    Outward (disc) & M\,51, M\,83,  &\cite{Fletcher+11}, \cite{Heald+16} and Fig.~\ref{fig:m83}\\
            & NGC\,891, NGC\,4013,   &\cite{Krause07,Stein+19b}\\
            & NGC\,4254, NGC\,4414,  & \cite{Chyzy08,Soida+02}\\
            & NGC\,4449 & \cite{Chyzy+00} (re-analyzed)\\
            & NGC\,4736, NGC\,5775 & \cite{Chyzy+Buta08,Soida+11}\\
    Inward (central region) \& outward (disc)   & -- & -- \\
    Inward (inner disc) \& outward (outer disc) & NGC\,4666 & \cite{Stein+19a}\\
    Outward (inner disc) \& inward (outer disc) & Milky Way & \cite{Brown+07}\\
    \hline
    \end{tabular}
    \end{center}
    \end{table*}

    \subsection{Parity of the Field}
    \label{sec:parity}
    Magnetic fields generated by mean-field dynamos in discs of galaxies tend to have quadrupole-like symmetry,
    with $\mbr(-z)=\mbr(z)$, $\mbp(-z)=\mbp(z)$ and $\mbz(-z)=-\mbz(z)$ (using cylindrical coordinates $r$, $\phi$, and $z$ with $z$ along the galactic rotation axis).
    Such configurations are also referred to as even parity or symmetric solutions. 
    The fastest growing eigenmode in the linear (i.e. exponentially growing) regime is found to be symmetric.
    In the non-linear regime, steady (saturated) symmetric solutions are obtained.
    These solutions are generally found to be non-oscillatory. 
    This is different from the odd parity, asymmetric, or dipole-like field configuration observed around the Sun, for instance.
    The solar field also oscillates with a $\sim22\yr$ period.
    Exponential growth rates of the various eigenmodes in 1-D dynamo solutions for a thin slab and a spherical shell
    were presented in Sec.~6 of \citet{Brandenburg+Subramanian05a}.
    Dipole-like symmetry is easier to excite in spherical objects, 
    and can be present in galactic halos, though possibly even in the disc under certain conditions \citep{Sokoloff+Shukurov90,Moss+Sokoloff08,Moss+10,Gressel+13b}; see Sec.~\ref{sec:halo_theory}.
    
    For $\alpha\Omega$ dynamos, large-scale magnetic field lines in the galactic disc near the midplane
    trace out spirals trailing  the galactic rotation.
    This results because the large-scale angular rotation speed decreases with radius in galaxies.
    Averaging the solenoidality condition on the field $\bfDel\cdot\bfB=0$, implies that the large-scale field must also be solenoidal $\bfDel\cdot\meanv{B}=0$.
    Thus, where the disc is locally thin such that the diffuse gas scale height $h\ll r$, 
    the mean field is almost parallel to the galactic midplane: $\mbz\ll\mbr$, $\mbp$.
    In some saturated dynamo solutions, field lines change from trailing to leading spirals 
    with $\mbr$ changing sign at a value of $|z|<h$ \citep[e.g.][]{Chamandy16}.
    
    The field parity with respect to the galaxy plane was investigated from $RM$ maps of a few edge-on galaxies (Table~\ref{tab:parity}), 
    showing clear preference for even parity.
    We will return to discuss halo magnetic fields in Sec.~\ref{sec:halo}.

    \subsection{Reversals of the Large-Scale Field}
    \label{sec:reversals}

    Reversals of the large-scale field are observed as reversals of the sign of Faraday rotation measures.
    Field reversals were observed between the central region and the disc of three spiral galaxies so far (Table~\ref{tab:directions}).
    Central regions are characterized by a rising rotation curve, i.e. weak differential rotation, which may be favourable for the $\alpha^2$ dynamo. Reversals within the disc were detected only in the Milky Way and in NGC\,4666, while in all other spiral galaxies observed so far the field direction remains the same within the disc (Table~\ref{tab:directions}).
    
    The seed mean field associated with non-zero averages over the small-scale field 
    (for finite scale separation; see Sec.~\ref{sec:theory_def}) will have both signs (Sec.~\ref{sec:seed}), 
    and hence different signs of the mean field can grow exponentially at different locations in the disc.
    Subsequently, the field smoothes out azimuthally so that reversals tend to form circles (or, presumably, cylinders) 
    of a given galactocentric radius, separating annular regions of oppositely signed mean field.
    If the random seed field is strong enough compared to the turbulence, 
    these global reversals can persist up to the non-linear regime \citep{Poezd+93} and  survive for several $\Gyr$
    (see \citep{Vasilyeva+94,Belyanin+94,Moss+12,Chamandy+13a,Moss+Sokoloff13,Moss+13} for detailed results and discussion). 
 
    \begin{table*}
    \begin{center}
    \caption{Parity of fields with respect to the midplane, as measured from maps of Faraday rotation in almost edge-on galaxies.}
    \label{tab:parity}
    \vspace{0.2cm}
    \begin{tabular}{@{}llc@{}}
    \hline
    Galaxy & Parity \& remarks & Reference \\
    \hline
    NGC\,253  & even    & \cite{Heesen+09b}\\
    NGC\,891  & even    & \cite{Krause09}\\
    NGC\,4013 & even (disc) + odd (central region) & \cite{Stein+19b}\\
    NGC\,4631 & even (disc) + varying (halo) & \cite{Mora+Krause13,Mora+19}\\
    NGC\,4666 & even    & \cite{Stein+19a}\\
    NGC\,5775 & even    & \cite{Soida+11}\\
    Milky Way & even (outer disc) + odd (central region) & \cite{Mao+12a,Schnitzeler+19}\\
    \hline
    \end{tabular}
    \end{center}
    \end{table*}

    Reversals have been found in various kinds of models.
    \citet{Gressel+13b} explored a more sophisticated mean-field model 
    that solves the averaged momentum equation in addition to the averaged induction equation.
    They found solutions with a superposition of even and odd parity axisymmetric modes, 
    which can produce reversals in the disc.
    \citet{Dobbs+16} performed an idealized galaxy simulation using smoothed particle MHD
    which included a steady, rigidly rotating spiral density wave as a component of the prescribed gravitational potential.
    They found that the non-axisymmetric velocity perturbations induced by the spiral can cause reversals in the magnetic field. 
    \citet{Pakmor+18} explored the magnetic field obtained in a disc galaxy from the Auriga cosmological MHD simulation,
    and found several reversals in the large-scale field at zero cosmological redshift.
    Here the reversals are between magnetic spiral arms (Sec.~\ref{sec:non-axisymmetric}),
    which suggests that the $m=0$ azimuthal component may not be dominant.
    Somewhat similarly, an apparent $m=2$ magnetic morphology 
    was obtained in a simulation of an isolated barred galaxy, where the magnetic field evolves
    according to the ``cosmic-ray driven dynamo'' model \citep{Kulpa-dybel+11}.
    \citet{Machida+13} ran a MHD DNS for an isolated galaxy and obtained a magnetic field 
    that has several reversals in radius at a given time in the saturated state, 
    but that also reverses quasi-periodically in time, with time scales of $\sim1.5\Gyr$.
    They interpreted their results as stemming from the interplay between the magneto-rotational and Parker instabilities.
    They obtain plasma $\beta$ of $\sim5$-$100$ in the saturated state, whereas observations suggest values $<1$ (Sec.~\ref{sec:strength_B_tot}).
    
    \subsection{Helicity of the Field}
    \label{sec:helicity}
    
    The importance of magnetic helicity in models (Sec.~\ref{sec:galactic_dynamo_theory}) motivates the observation of helicity.
    Magnetic helicity and its volume density are in general gauge dependent quantities \citep{Brandenburg+Subramanian05a}, 
    and hence studies relating to observations have focussed on the closely related quantity, the current helicity density 
    $\propto\bfB\cdot\bfDel\cro\bfB$, which is a measure of the helical twisting of fields lines.%
    \footnote{A gauge independent formulation of the magnetic helicity density exists 
    for the small-scale random component of the field \citep{Subramanian+Brandenburg06}.}
    When observing a helical field with the axis of the helix aligned with the line of sight, for example, 
    the rotation of the field leads to a rotation of the polarization plane
    that is either in the same sense as that produced by the Faraday rotation,
    or in the opposite sense, depending on the relative signs of helicity and Faraday depth.
    In the first case, this leads to extra rotation and
    depolarization in addition to that produced by Faraday rotation.
    In the second case, the helicity and the Faraday rotation
    counteract and partly cancel one another.
    This could even lead to ``anomalous depolarization'',
    where the degree of polarization 
    actually increases with increasing wavelength (see Fig.~9 in \citep{Sokoloff+98}).
    More generally, one could, in principle, measure statistical correlations between polarization fraction and Faraday depth
    and use these to probe the helicity of the magnetic field.
    The potential of these sorts of methods has been demonstrated using idealized models and simulations,
    but the feasibility of applying them to real data is still unclear \citep{Volegova+Stepanov10,Brandenburg+Stepanov14,Horellou+Fletcher14}.

    \subsection{Boundary Conditions in Mean-Field Models}
    \label{sec:drivers_dynamo}
    
    Dynamo modelers often assume vacuum electromagnetic boundary conditions on $\meanv{B}$  \citep[][VII.5]{Ruzmaikin+88}
    for $|z|>h$ and $r>R$, where $h$ is the galactic scale height of diffuse gas and $R$ its scale length.
    The quantity $h$ is best estimated as the scale height of the warm phase of the ISM, 
    as local DNSs of the supernova (SN)-driven turbulent ISM find that the saturated large-scale field resides primarily in the warm phase, 
    not the transient hot phase or cold phase that has a small volume filling fraction \citep{Evirgen+17}.
    
    The magnetic field evolution is not expected to be sensitive to the degree of ionization 
    because ambipolar diffusion strongly couples ionized and neutral gas even for the low ionization fractions
    $\sim0.007$-$0.08$ \citep{Kulkarni+Heiles87,Jenkins+13} estimated for the diffuse warm neutral medium.
    This can be seen by computing the mean time for a given neutral to collide with an ion
    and comparing it to other relevant time scales \citep[e.g.][]{Blaes+Balbus94}.
    We estimate this timescale to be $\lesssim10^4\yr$ in the disc,
    which is small compared to the turbulent correlation time ($\sim10^7\yr$) and galactic rotation period.
    Hence, the diffuse warm gas can be treated as a single entity in dynamo models,
    and we do not consider the effects of ambipolar diffusion on the magnetic field \citep[e.g.][]{Brandenburg19}.
    
    If the large-scale field is assumed to be axisymmetric, vacuum boundaries imply $\mbp=0$ everywhere outside the disc.
    If $h\ll R$, then also $\mbr\approx0$ on the disc surfaces. 
    So vacuum boundary conditions imply $\mbr\simeq\mbp=0$ at $z=\pm h$.
    Real galaxies can be non-axisymmetric and include halos and outflows from the disc into the halo 
    and there remains work to be done to explore dynamo models for more general boundary conditions.
    
    \section{Strength of the Magnetic Field}
    \label{sec:strength}
    
    \subsection{Total Field Strength}
    \label{sec:strength_B_tot}
    
    The strength of the \textit{total} magnetic field $B_\mathrm{tot}$ has been typically derived from the measured synchrotron intensity $I$ by assuming energy density equipartition between total magnetic fields and total cosmic rays (CRs) \citep{BK05}.
    The assumption of CR equipartition  may be less robust than other aspects of magnetic field observations.
    Empirically, \citet{Seta+Beck19} found that this is
    valid for star-forming spiral galaxies at scales above about 1\,kpc, but probably not on smaller scales and not for galaxies experiencing a massive starburst or for dwarf galaxies.
    
    To improve the interpretation of galactic synchrotron emission, a better understanding of the spatial distribution of cosmic rays in galaxies is needed.
    In this respect we note that global MHD galaxy simulations are now capable of modeling CR transport. 
    For example, \citet{Pakmor+16} found that by setting the CR diffusivity to be non-zero only  parallel to the magnetic field (anisotropic diffusion),
    CRs become confined to the galactic disc, whereas they mostly diffuse out of the disc if this same diffusivity is assumed in all directions (isotropic diffusion).
    However, given that the Larmor radius is several orders of magnitude below the resolution of such simulations,
    more sophisticated prescriptions for CR transport based partly on test particle simulations \citep[e.g.][]{Snodin+16,Rodrigues+19b}
    may ultimately prove to be necessary.
    
    According to \citet{Fletcher10}, the typical total field strength given in the recent literature for 21 bright spiral galaxies is $B_\mathrm{tot}\approx17\,\mu$G. The energy density of the total (mostly unresolved turbulent) magnetic field is found to be comparable to that of turbulent gas motions and about one order of magnitude larger than the energy density of the thermal gas \citep{Beck07,Beck15b,Tabatabaei+08}.
    This implies that the turbulence is supersonic.
    Can this result be reconciled with previous work which suggests that turbulence in warm gas is trans-sonic \citep[e.g.][]{Hill+08,Gaensler+11,Iacobelli+14} ?
    
    Supersonic turbulence is expected to support a saturated fluctuating magnetic field 
    with considerably smaller energy density than turbulent motions.
    This is because supersonic (as opposed to subsonic) turbulence has a more direct non-local transfer 
    of energy to dissipation scales, 
    where some of the available energy that could otherwise be used for amplification just turns into heat.
    This idea is consistent with results of small-scale dynamo DNSs by \citet{Federrath+11} 
    which suggest that the saturated small-scale magnetic field is relatively small 
    compared to $B\eq$ when the turbulence is supersonic.
    On the other hand, those simulations do not include a large-scale dynamo, which could lead to a larger total field strength.
    Furthermore, as noted by \citet{Kim+Ostriker15b}, the simulations are not run up until the full saturation of the field growth.
    
    Another explanation for this discrepancy may be that the generic assumption of CR equipartition with the magnetic field is simply inexact,  and the field could be lower than CR equipartition  would imply.  Even though CR equipartition might be the minimum energy relaxed state, galaxies are being forced  with some combination of turbulence and CRs.
    The steady state may more accurately be characterized as an intermediate equilibrium state, balanced by forcing against, and relaxation toward, the relaxed state.
    
    In Table~\ref{tab:strength} we compile data on the strengths of the various magnetic field components in radial rings for 12 galaxies.
    Instrumental noise in $I$ hardly affects $B_\mathrm{tot}$, so that the values in Table~\ref{tab:strength} have small rms errors. 
    The main uncertainties are due to those in pathlength through the synchrotron-emitting disc of $L_\mathrm{syn}/\cos(i)$ (where $i$ is the inclination against face-on view) and the proton/electron ratio $K$ \cite{Seta+Beck19}. We assumed $L_\mathrm{syn}=1$\,kpc (except for M\,31) and a constant CR proton to electron ratio $K=100$; an uncertainty by a factor of two in either quantity amounts to a systematic error in $B_\mathrm{tot}$ of about 20\%. For those galaxies in Table~\ref{tab:strength} for which only total intensities are available, we subtracted the average thermal fraction of 20\% at 4.86\,GHz \cite{Tabatabaei+17} or 30\%, extrapolated to 8.46\,GHz. An uncertainty in thermal fraction by a factor of two causes a negligible error in $B_\mathrm{tot}$ of about 5\%. A distance uncertainty does not affect $B_\mathrm{tot}$ because synchrotron intensity (surface brightness) is independent of distance. The uncertainty in inclination $i$ affects $B_\mathrm{tot}$ via the dependence on pathlength; a typical uncertainty of $\pm3^\circ$ in $i$ leads to less than 15\% uncertainty in pathlength for galaxies inclined by $i\le70^\circ$ and to an even smaller uncertainty in $B_\mathrm{tot}$. The strength of $B_\mathrm{tot}$ (averaged within each galaxy and then between galaxies) is $13\,\mu$G with a dispersion of $4\,\mu$G between galaxies.
    
    \subsection{Ordered Field Strength}
    \label{sec:strength_B_ord}
    
    The strength of the \textit{ordered} field $B_\mathrm{ord}$ is derived from the strength of the total field and the degree of synchrotron polarization $P$. The uncertainty of $B_\mathrm{ord}$ is the same as for the total field of about 20\%; the uncertainty in $P$ does not contribute significantly because only averages within radial rings are considered in Table~\ref{tab:strength}.
    $P$ and hence $B_\mathrm{ord}$ may increase with higher spatial resolution because the field structure can be better resolved. The nearby galaxies M\,31 and M\,33 were observed with $3'-5'$ angular resolution, while the more distant galaxies (see Table~\ref{tab:pitch}) were observed with about $15''$ (except M\,101), yielding roughly similar spatial resolutions of $0.4-1.4$\,kpc.
    
    According to Table~\ref{tab:strength}, the strength of $B_\mathrm{ord}$ (averaged within each galaxy and then between galaxies) is $5\,\mu$G with a dispersion of $2\,\mu$G between galaxies.
    Taking the averages for each galaxy, $B_\mathrm{ord}$ is $1.4-4.2\times$ smaller than $B_\mathrm{tot}$. 
    This means that the strength of the observationally unresolved field $B_\mathrm{iso}$
    dominates and comprises typically $75-97\%$ of the total field strength.
    While $B_\mathrm{tot}$ decreases radially in almost all galaxies, $B_\mathrm{ord}$ remains about constant.

    \begin{table}
        \caption{Strengths of the total field $B_\mathrm{tot}$ (derived from the radio synchrotron intensity at 4.86\,GHz, averaged in radial rings in the galaxy plane, assuming energy equipartition between CRs and magnetic fields, a proton/electron ratio of 100, and a thickness of the synchrotron-emitting disc of 1\,kpc for all galaxies except M\,31), of the large-scale regular field $B_\mathrm{reg}$ (derived from the amplitude $|RM_0|$ of the axisymmetric azimuthal mode ($m=0$) or of the bisymmetric mode ($m=1$) in the radial ring, the inclination $i$, a thickness of the thermal disc of 1.4\,kpc for all galaxies except M\,31, and the average thermal electron density $\langle n_\mathrm{e}\rangle$, scaled according to the total field strength $B_\mathrm{tot}$), and of the ordered field $B_\mathrm{ord}$ (derived from $B_\mathrm{tot}$ and the degree of synchrotron polarization at 4.86\,GHz). The last column gives the reference for the $|RM_0|$ measurements.
        Galaxy distances are given in Table~\ref{tab:pitch}.}
    \label{tab:strength}
    \vspace{0.2cm}
    \centering
    \begin{tabular}{l l l l l l l l l l l}
    \hline
    Galaxy & Radial range &	$|RM_0|$ & $i$ & $B_\mathrm{tot}$ & $\langle n_\mathrm{e}\rangle$ & $B_\mathrm{reg}$ &  $B_\mathrm{ord}$ & $B_\mathrm{reg}/$ & $B_\mathrm{ord}/$ & Ref.\\
           & [kpc]   & [rad/m$^2$] & [$^{\circ}$] & [$\mu$G] & [cm$^{-3}$] & [$\mu$G] & [$\mu$G] & $\,B_\mathrm{tot}$ & $\,B_\mathrm{reg}$ & \\
    \hline
    M\,31  & 6.8--9.0 $^a$  & $83\pm7$ & 75  & 7.3 $^b$ & 0.032 & 1.8 $^c$ & 4.9  &        0.25 & 2.7 & \cite{Fletcher+04}\\
           & 9.0--11.3 $^a$ & $96\pm9$ &     & 7.5 $^b$ & 0.033 & 2.1 $^c$ & 5.2  & 0.28 & 2.5 \\
           & 11.3--13.6 $^a$ & $115\pm9$ &   & 7.1 $^b$ & 0.031 & 2.6 $^c$ & 4.9  & 0.37 & 1.9 \\
           & 13.6--15.8 $^a$ & $99\pm6$ &    & 6.3 $^b$ & 0.026 & 2.7 $^c$ & 4.6  & 0.43 & 1.7 \\
    M\,33  & 1.0--3.0   & $69\pm4$     & 56  & 8.7 & 0.031  & 1.3 & 3.1  & 0.15 &
           2.4 & \cite{Tabatabaei+08}\\
           & 3.0--5.0   & $103\pm9$    &     & 7.6 & 0.026  & 2.4 & 3.1  & 0.32 & 1.3 \\
    M\,51  & 2.4--3.6   & $46\pm3$     & 20  & 17  & 0.086  & 1.3 & 8.6  & 0.08 &
           6.6 & \cite{Fletcher+11}\\
           & 3.6--4.8   & $57\pm15$    &     & 16  & 0.078  & 1.8 & 7.6  & 0.11 & 4.2 \\
           & 4.8--6.0   & $76\pm21$    &     & 15  & 0.071  & 2.6 & 7.6  & 0.17 & 2.9 \\
           & 6.0--7.2   & $76\pm2$     &     & 13  & 0.057  & 3.2 & 7.8  & 0.25 & 2.4 \\
    M\,81  & 6.0--9.0  & $20\pm4$ $^{d,\,e}$& 59  & 8.0 & 0.028 & 0.4 $^e$
           & 4.1 $^f$ & -- & -- & \cite{Krause+89b}\\
           & 9.0--12.0 & $20\pm3$ $^{d,\,e}$&     & 6.4 & 0.020 & 0.5 $^e$
           & 3.8 $^f$ & -- & -- &\\
    M\,83  & 4--8   & $27\pm13$ $^{d}$ & 24  & 19  & 0.101  & 0.5 & 5.9 
           & 0.03 & 12 & Table~\ref{tab:m83}\\
           & 8--12  & $83\pm19$ $^{d}$ &     & 16  & 0.078  & 2.1 & 6.5 
           & 0.13 & 3.1 & \\
    NGC\,253  & 1.4--6.7 & $120\pm20$  & 78.5& 15  & 0.071  & \textit{0.30} $^g$
           & 4.3 & \textit{0.02} & \textit{14} & \cite{Heesen+09b}\\
    NGC\,1097 & 3.75--5.0 $^h$ & $155\pm8$  & 45  & 13  & 0.057  & 2.4 & 7.9  & 0.18 &
              3.3 & \cite{Beck+05}\\
    NGC\,1365 & 2.625--4.375 & $55\pm3$ & 40 & 15  & 0.071  & 0.8 & 4.8  & 0.05 &
              6.0 & \cite{Beck+05}\\
              & 4.375--6.125 & $65\pm2$ &    & 11  & 0.045  & 1.5 & 5.7  & 0.14 & 3.8 \\
              & 6.125--7.875 & $52\pm6$ &    & 12  & 0.051  & 1.1 & 4.7  & 0.09 & 4.3 \\
              & 7.875--9.625 & $100\pm1$ &   & 13  & 0.057  & 1.8 & 4.0  & 0.14 & 2.2 \\
              & 9.625--11.375 & $90\pm10$ &  & 12  & 0.051  & 1.8 & 3.8  & 0.15 & 2.1 \\
              & 11.375--13.125 & $56\pm14$ & & 10  & 0.039  & 1.5 & 3.4  & 0.15 & 2.3 \\
              & 13.125--14.875 & $32\pm6$ &  & 8.4 & 0.030  & 1.1 & 2.9  & 0.13 & 2.6 \\
    NGC\,4254 & 4.8--6.0& $68\pm12$  $^i$ & 42  & 18  & 0.094  & 0.7 & 7.8  & 0.04 &
              11 & \cite{Chyzy08}\\
              & 6.0--7.2& $87\pm9$  $^i$  &     & 17  & 0.086  & 1.0 & 8.7  & 0.06 & 8.7 \\
              
    NGC\,4449 & 1.0--2.0 & $62\pm12$ $^i$  & 43  & 16  & 0.078  & 0.7 & 4.1  & 0.04           & 5.9 & \cite{Chyzy+00}\\
              & 2.0--3.0 & $67\pm14$ $^i$  &     & 11  & 0.045  & 1.4 & 4.8  & 0.13 & 3.4 \\
    NGC\,6946 & 0--4.7   & $81\pm8$ $^j$ &30 & 19  & 0.101  & 1.2 & 5.1  & 0.06 &
              4.3 & \cite{Ehle+Beck93}\\
              & 4.7--9.4 & $72\pm11$ $^j$ &  & 13  & 0.057  & 1.9 & 5.1  & 0.15 & 2.7\\
    IC\,342   & 7.5-12.5 & $8\pm2$     & 31  & 14  & 0.064  & \textit{0.18} $^g$
              & 3.3 & \textit{0.013} & \textit{18} & \cite{Beck15a}\\
              & 12.5-17.5& $6\pm2$     &     & 12  & 0.051  & \textit{0.17} $^g$
              & 2.9 & \textit{0.014} & \textit{17} &\\
    \hline
    \end{tabular}
    \vspace{0.2cm}
    \footnotesize
    \begin{itemize}
    \item[] $^a$ Scaled to 780\,kpc distance; $^b$ assuming a full thickness of the synchrotron-emitting disc of 440--580\,pc \cite{Fletcher+04}; $^c$ assuming a full thickness of the thermal disc of 500\,pc \cite{Walterbos+Braun94}; $^d$ amplitude of the $m=1$ mode; $^e$ lower limit due to Faraday depolarization at 1.4\,GHz; $^f$ derived from 4.85\,GHz data \cite{Beck+85} that are less affected by Faraday depolarization; $^g$ exceptionally weak regular field; $^h$ the other radial ranges do not allow a continuous sinusoidal fit; $^i$ re-analyzed from the original data; $^j$ amplitude of the mode superposition $m=0+2$.
    \end{itemize}
    \normalsize
    \end{table}
    
\begin{figure*}[t]
    \vspace*{7mm}
    \begin{center}
    \includegraphics[width=14cm]{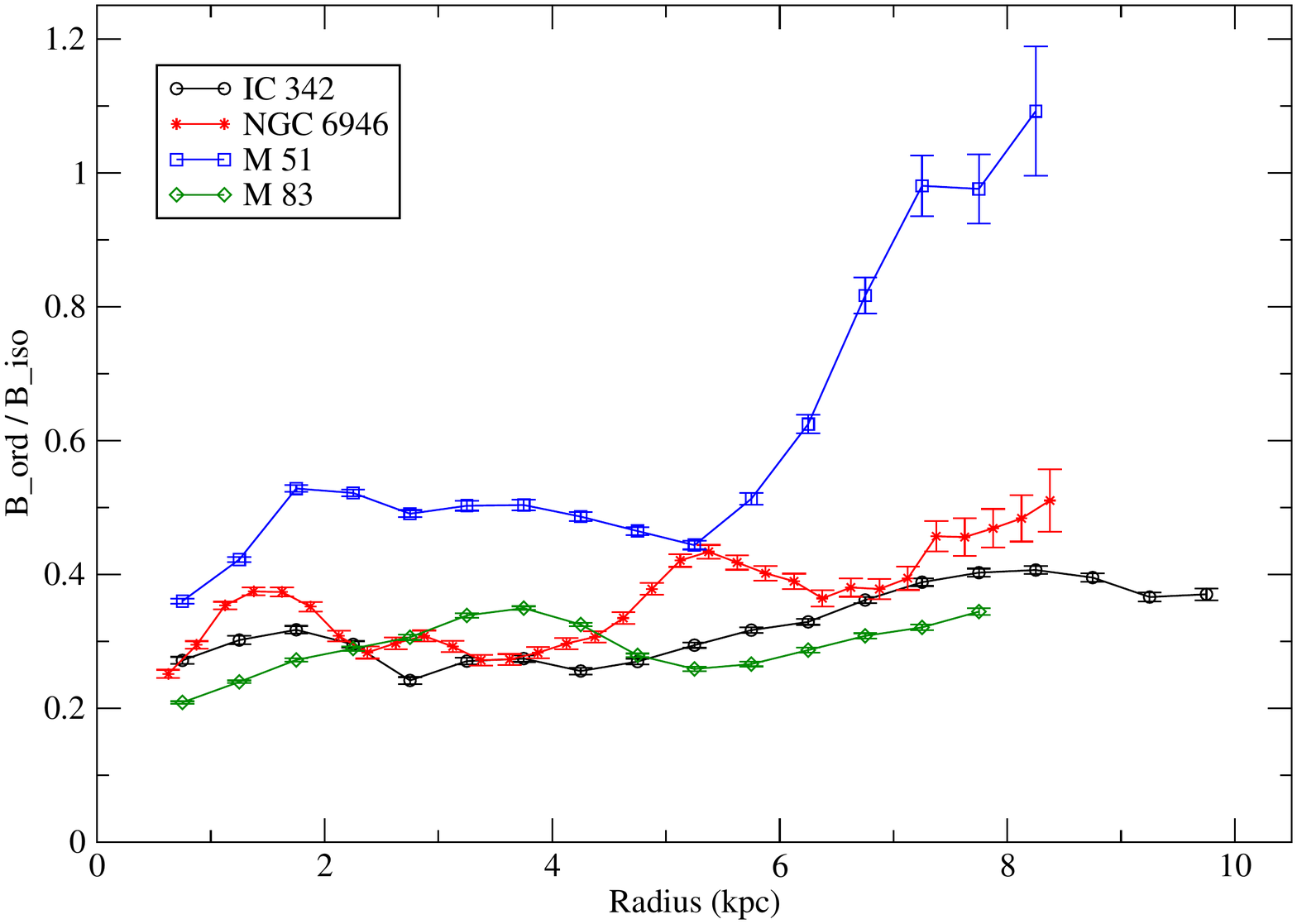}
    \caption{Radial variation of the ratio of ordered to unresolved (isotropic turbulent) field strengths $q\equiv B_\mathrm{ord}/B_\mathrm{iso}$ , derived from the observed degree of polarization of the synchrotron emission at 4.86\,GHz (6.2\,cm wavelength), averaged over all azimuthal angles of each radial ring in the galaxy's plane (i.e. corrected for inclination).
    The largest radius of each plot is limited by the extent of the map of thermal emission that needs to be subtracted from the map of total emission to derive the map of total synchrotron emission.}
    \label{fig:ratio}
    \end{center}
    \end{figure*}

    The ratio $q$ between the strengths of the ordered field $B_\mathrm{ord}$ and the unresolved field $B_\mathrm{iso}$ can be directly computed from the observed degree of polarization of the synchrotron emission $P$, using Eq.~(4) in \citet{Beck07} under the assumption of equipartition between magnetic fields and CRs. 
    The uncertainty of $q$ is given by the observational uncertainties in $P$ only and is not affected by the uncertainties in the estimates of the absolute values of field strengths.  Figure~\ref{fig:ratio} shows the result for four of the best-studied spiral galaxies.
    $q$ generally increases with increasing distance from a galaxy's centre.
    The magnetic arms of NGC\,6946 and M\,83 (see Sect.~\ref{sec:non-axisymmetric}) are prominent as regions with high $q$, i.e. a highly ordered field, whereas the optical arms generally reveal low values of $q$. The magnetic field in the outer parts of M\,51 is exceptionally ordered, probably due to shearing gas flows caused by the gravitational interaction with the companion galaxy.
    
    \subsection{Regular Field Strength}
    \label{sec:strength_B_reg}
    
    Measuring the strength of the \textit{regular} field $B_\mathrm{reg}$ requires  Faraday rotation measures, $RM$ derived from diffuse polarized synchrotron emission at high frequencies (where Faraday depolarization is small), and knowledge about the inclination angle between $B_\mathrm{reg}$ and the sky plane, the average density $\langle n_\mathrm{e}\rangle$ of the thermal electrons, and the pathlength $L_\mathrm{th}$ through the disc of thermal gas.
    Ideal tools are $RM$ and dispersion measure $DM$ measurements of pulsars, but are available only in the Milky Way \citep[e.g.][]{Han+06}
    \footnote{$RM$ and $DM$ was measured for just one pulsar in the Large Magellanic Cloud; many more are needed to estimate $B_\mathrm{reg}$ in this galaxy.}.
    $|RM_0|$ is the measured amplitude of the sinusoidal azimuthal mode in a radial ring, so that $B_\mathrm{reg}$ is the axisymmetric regular field on the scale of the galaxy (see Sec.~\ref{sec:non-axisymmetric}).
    
    As very little is known about $L_\mathrm{th}$ in external galaxies, we assume that the exponential scale height $H_\mathrm{th}$ of the thermal ``thick disc'' is the same as in our Milky Way, $H_\mathrm{th}\simeq1.0$\,kpc \cite{Berkhuijsen+Mueller08,Savage+Wakker09}
    \footnote{Yao et al. \cite{Yao+17} described the Milky Way's ``thick disc'' of thermal gas by a $sech^2$ function with a scale height of $1.67$\,kpc. The exponential tail of this function has an exponential scale height of 1.0\,kpc.}
    , and that it is constant at all radii
    \footnote{Models of the disc of warm neutral gas, based on \hi\ data of edge-on galaxies, indicate that the full thickness to half maximum increases beyond about 5\,kpc radius (``flaring'') \cite{Peters+17}, while the radial dependence of the disc thickness of ionized gas has not been investigated so far.}.
    The full thickness at half maximum is $2\, ln2\, H_\mathrm{th}\simeq1.4\, H_\mathrm{th}$, so that we use $L_\mathrm{th}=1.4\,\mathrm{kpc}/\cos(i)$, except for M\,31
    \footnote{The thermal gas in M\,31 is distributed in an ellipsoidal ring in the plane of the sky with a Gaussian profile of about 6\,kpc full width to half maximum (Berkhuijsen, priv. comm.) and 200--500\,pc exponential scale height \cite{Walterbos+Braun94}, corresponding to about 500\,pc full thickness, which yield a pathlength of $L_\mathrm{th}\simeq1.8\,$kpc for 75$^{\circ}$ inclination.}.

    In principle, $\langle n_\mathrm{e}\rangle$ can be measured from the intensity of thermal emission in the radio or optical ranges; however, thermal intensity is proportional to $\langle n_\mathrm{e}^2\rangle$, while $RM$ is proportional to $\langle n_\mathrm{e}\rangle$, so that knowledge of the filling factor $f =\, \langle n_\mathrm{e}\rangle^2 / \langle n_\mathrm{e}^2\rangle$ is needed, but lacking for external spiral galaxies.
    Furthermore, the pathlengths of thermal emission and thermal gas density are different.
    Instead, we assume that $\langle n_\mathrm{e}\rangle^2$ is linearly related to the total star-formation rate ($SFR$) in a galaxy, as is indicated by the similarity of the various $SFR$ tracers \cite{Tabatabaei+17}. Radio synchrotron intensity varies non-linearly with $SFR$, which can be expressed as $B_\mathrm{tot} \propto SFR^{\,0.34\pm0.04}$ \cite{Tabatabaei+17} \footnote{M\,31 deviates from this relation. For the average $B_\mathrm{tot}=7.0\,\mu$G in the radial range 7--16\,kpc (Table~\ref{tab:strength}), the above equation predicts $SFR\approx0.17$\,M$_\odot$/yr, while $SFR\approx0.3$\,M$_\odot$/yr is measured in this radial range \cite{Tabatabaei+10}. Consequently, the values $\langle n_\mathrm{e}\rangle$ in Table~\ref{tab:strength} are increased by a factor of $\sqrt{0.3/0.17}$.}
    , in agreement with the theoretical expectation of $B_\mathrm{tot} \propto SFR^{1/3}$ if turbulence is driven by supernovae and the energy of the turbulent magnetic field is a fixed fraction of the turbulent energy \cite{Schleicher+Beck13, Schober+16}.
    The combination gives $\langle n_\mathrm{e}\rangle \propto \langle B_\mathrm{tot}\rangle^{\approx1.5}$ (assuming a constant filling factor $f$). This relation is calibrated with data from the Milky Way where $\langle n_\mathrm{e}\rangle \approx 0.018$\,cm$^{-3}$, which is about constant along the sightlines to pulsars located mostly within about 2\,kpc from the Sun \cite{Berkhuijsen+Mueller08}, and an equipartition strength of the local total field of $B_\mathrm{tot} \approx 6$\,$\mu$G \cite{Seta+Beck19}. The estimates of $\langle n_\mathrm{e}\rangle$ are given in column 6 of Table~\ref{tab:strength}. The values for M\,51 agree well with the estimates derived from thermal radio emission \cite{Berkhuijsen+97}.
    
    Following the above procedure, the strengths of $B_\mathrm{reg}$ are re-computed and listed in Table~\ref{tab:strength}. (Note that many ``$\bar{B}$'' values in Table~2 of \citet{Vaneck+15} refer to the ordered, not the regular field.)
    As $B_\mathrm{reg}$ is derived from the amplitude of the large-scale $RM$ variation and all galaxies in Table~\ref{tab:strength} were observed with sufficiently high resolution, $B_\mathrm{reg}$ does not depend on the actual resolution. The uncertainty in $B_\mathrm{reg}$ is quite large but difficult to quantify; the uncertainties of the assumptions lead to an uncertainty of roughly 30--40\%, while the relative uncertainties of the $RM$ observations are generally smaller.
    
    The range of variation of $B_\mathrm{reg}$ is larger than that of the total field. 
    Some galaxies like M\,31, M\,33, and M\,51 reveal strengths of as large as $10-30\%$ of the total field, 
    while $B_\mathrm{reg}$ in IC\,342 is weaker than 2\% of the total field strength.
    The average strength of the regular field $B_\mathrm{reg}$ is $1.7\,\mu$G, with a dispersion of $0.6\,\mu$G between galaxies (excluding the exceptionally weak regular fields of the galaxies NGC\,253 and IC\,342),
    similar to that of the local Milky Way of $1.5\pm0.2\,\mu$G derived from pulsar data \cite{Rand+89,Han+94}.
    The average ratio $B_\mathrm{reg}/B_\mathrm{tot}$ is $0.14\pm0.09$.
    $B_\mathrm{reg}$ increases radially in most galaxies.
    An increasing ratio $B_\mathrm{reg}/B_\mathrm{tot}$ indicates that the large-scale dynamo becomes more efficient 
    relative to the small-scale dynamo at larger radii.
    Pulsar data from the Milky Way seem to indicate an opposite trend (see Fig.~11 in \cite{Han+06}). However, no clear azimuthal modes could be identified in the Milky Way \cite{Men+08}, and $B_\mathrm{reg}$ refers to averages over sight lines to individual pulsars, tracing smaller spatial scales.
    
    The ordered field $B_\mathrm{ord}$ in Table~\ref{tab:strength} is always larger than the regular field $B_\mathrm{reg}$. The average ratio $B_\mathrm{ord}/B_\mathrm{reg}$ is 4.0 with a dispersion of 2.4 (again excluding NGC\,253 and IC\,342). One reason for this high ratio is that most values of $B_\mathrm{reg}$ (except that for NGC\,6946) refer to the large-scale axisymmetric azimuthal mode, i.e. the galaxy-wide regular field. $RM$ patterns on smaller scales are observed in all galaxies and are signatures of structures of regular fields on many scales. This may also explain why $B_\mathrm{reg}$ measured from pulsar $RMs$ possibly increases toward the inner Milky Way. Furthermore, anisotropic turbulent fields contribute to $B_\mathrm{ord}$ but not to $B_\mathrm{reg}$.
        
    \subsection{Mean-Field Strength from Dynamo Models}
    \label{sec:strength_theory}
    
    The saturated dynamo solutions are most relevant for comparison with observation.
    This is because large-scale fields in galaxies are generally of order $\mkG$ strength (Table~\ref{tab:strength}), 
    which is similar to $B\eq$ (equation~\ref{Beq}), so the dynamo would be in the non-linear regime.
    Moreover, mean-field growth is fairly rapid with
    global eigenmodes in the kinematic regime that have 
    e-folding times of $t_\mathrm{e}=1/\Gamma\sim(1-5)\times10^8\yr$,
    where $\Gamma$ is the global growth rate \citep[e.g.][]{Chamandy16}.
    The saturated state will itself evolve as the underlying galactic parameters evolve.
    
    \citet{Rodrigues+15} computed the evolution of the magnetic fields of a large sample of galaxies over cosmic time.
    For each galaxy at a given time, they adopted a steady-state mean-field dynamo solution 
    depending on the underlying parameter values at that time.
    This is equivalent to assuming that the magnetic field adjusts instantaneously to changes in the underlying parameters.
    This assumption is relaxed in the subsequent model of \citet{Rodrigues+19a},
    where the dynamo equations are time-dependent, and thus the assumption can be checked.
    It is found that at any given time, model galaxies that
    have a non-negligible mean field typically also have dynamos that are operating close to critical (negligible growth or decay),
    while those that have negligible mean-field have subcritical dynamos.
    Moreover, the two models agree qualitatively.
    Hence, dynamo timescales are generally sufficiently small compared to  galactic evolution timescales
    to approximate the mean magnetic field as a (slowly evolving) steady-state dynamo solution.
    Major mergers are an exception to this rule, because during these events the underlying parameters can change rapidly.
    
    In the saturated nonlinear state, 
    solving only the \textit{local} mean-field dynamo problem (1-D in $z$)
    by neglecting terms involving radial and azimuthal derivatives 
    (the slab approximation) allows constructing axisymmetric \textit{global} solutions
    (2-D in $r$ and $z$) by stitching together local solutions along the radial coordinate. 
    These solutions turn out remarkably similar to fully global solutions for which radial derivative terms are not neglected \citep{Chamandy16}.
    
    Moreover, by employing the 'no-$z$' approximation to replace $z$-derivatives by divisions by $h$ with suitable numerical coefficients \citep{Subramanian+Mestel93,Moss95,Phillips01,Sur+07b,Chamandy+14b},
    and neglecting the $\alpha^2$ effect,
    one can write down an analytical solution for the large-scale field strength. 
    The resulting solutions are weighted spatial averages over $z$,
    but the vertical dependence can be reconstructed using perturbation theory along with $\Del\cdot\meanv{B}=0$
    (see Refs.~\citep{Chamandy+14b,Chamandy16} for details).
    The no-$z$ solution is given by
    \begin{equation}
      \label{Bsat}
      \mean{B}\simeq B\eq\left(\frac{D}{D\crit} -1\right)^{1/2}\frac{l}{h}\left(R_U +\pi^2 R_\kappa\right)^{1/2},
    \end{equation}
    where $D>D\crit$ (supercritical dynamo) has been assumed, $l$ is the correlation length of the turbulence, 
    $B\eq$ is given by equation~\eqref{Beq},
    the dynamo number is given by
    \begin{equation}
      \label{D}
      D\simeq -9q\shear\left(\frac{h\Omega}{u}\right)^2,
    \end{equation}
    with $q\shear=-d\ln\Omega/d\ln r$ the shear parameter, 
    the critical dynamo number is given by
    \begin{equation}
      \label{Dcrit}
      D\crit\simeq -\frac{\pi^5}{32}\left(1 +\frac{1}{\pi^2}R_U\right)^2,
    \end{equation}
    $R_U\equiv U\f h/\etat$ is the turbulent Reynolds number for the large-scale vertical outflow with characteristic speed $U\f$,
    the turbulent diffusivity is given by
    \begin{equation}
      \label{etat}
      \etat\simeq \frac{1}{3}\tau u^2,
    \end{equation}
    with $\tau\simeq l/u$ the turbulent correlation time,
    and the quantity $R_\kappa$ (see below) 
    is of order unity.
    Equation~\eqref{D} assumes that in the kinematic regime, $\alpha\simeq\tau^2u^2\Omega/h$
    \citep{Krause+Radler80},
    but this  applies if $\Omega\tau<1$ and $\alpha<u$;
    otherwise the expression should be modified \citep{Ruzmaikin+88,Chamandy+16,Zhou+Blackman17}.
    Equation~\eqref{Bsat} results from solving the mean induction equation along with the dynamical $\alpha$-quenching formalism
    (Sec.~\ref{sec:galactic_dynamo_theory}), and assumes that the $\Omega$ effect is strong compared to the $\alpha$ effect.
    
    The above analytic solution can depend parametrically on $r$.
    The scale height of diffuse gas $h$ increases with $r$ in a flared disc. 
    The root-mean-square turbulent speed $u$ might decrease with $r$
    (see Sec.~\ref{sec:velocity_dispersion_radial}),
    while for a flat rotation curve $q\shear=1$ and $\Omega\propto 1/r$.
    How the turbulent scale $l$, correlation time $\tau$, and outflow speed $U\f$ vary with radius is not clear.
    
    The field strength in equation~\eqref{Bsat} depends on $R_U$ in two competing ways:
    (i) $|D\crit|$ increases with $R_U$ leading to a less supercritical dynamo and thus a lower saturation strength;  
    (ii) a stronger outflow enhances the advective flux of small-scale magnetic helicity, 
    which helps to avert catastrophic quenching and favours a higher saturated field strength.
    The term containing $R_\kappa$ accounts for a turbulent diffusive flux of magnetic helicity \citep{Kleeorin+02} 
    and has also been measured in simulations \citep{Mitra+10}.
    If diffusive flux of magnetic helicity is present,
    the net effect of outflows is usually to hamper the dynamo, 
    but there exists a region of parameter space (at large $D/D\crit$, small $R_\kappa$, and small $R_U$) 
    for which outflows have a net positive effect on the saturated value of $\mean{B}$.
    While the advective flux has been derived from first principles \citep{Subramanian+Brandenburg06}, 
    such a derivation has not yet been carried out for the diffusive flux,%
    \footnote{Other helicity flux terms might also play a role \citep{Subramanian+Brandenburg06}.}
    but outflows are not required to explain the existence of 
    near-equipartition strength large-scale fields in galaxies if diffusive fluxes are present.
    
    The above solution is rather crude and its parameter values are often not well constrained,
    but it can be useful for making simple estimates.
    Further work involving detailed comparison of models with the data presented in this work is warranted.
    
    \section{Seed Fields and Small-Scale Fields in Mean-Field Dynamos}
    \label{sec:seed_ss}
    
    \subsection{Seed Fields}
    \label{sec:seed}
    
    As $\meanv{B}=0$ is a valid solution of the averaged induction equation in mean-field dynamo theory, 
    a finite seed mean field is required for dynamo amplification.
    There are many proposed mechanisms to generate a seed field in the early Universe 
    or during galaxy formation \citep{Subramanian16}, 
    but the resulting seed fields tend to be too small for typical mean field dynamo amplification on a galactic rotation to provide enough e-foldings in the available time to explain observed regular field strengths.
    A much stronger and more promising seed for the mean field,
    of $\sim10^{-4}B\eq$, can be supplied by the saturated fluctuating field 
    arising from the fast-acting fluctuation or small-scale dynamo (Sec.~\ref{sec:ss}).
    The fluctuation dynamo generates a field that peaks on small scales, but also generates low-level random large-scale \textit{perturbations} 
    which then seed the mean-field or large-scale dynamo
    \citep[Ch.~VII.14 of][]{Ruzmaikin+88} and Refs.~\citep{Beck+94,Subramanian+Brandenburg14}.
    Such seed fields are sufficiently strong for mean-field dynamo growth
    to explain large-scale field strengths inferred from observations, 
    even in some high-redshift galaxies \citep{Mao+17} \citep[see, e.g.][for models]{Arshakian+09,Rodrigues+19a}.
    
    However, many models implicitly assume that ``mean'' implies the ``ensemble mean'',
    so that averages represent infinite ensemble averages, and hence fluctuating fields average to precisely zero
    (Sec.~\ref{sec:theory_def}).
    Strictly speaking, these assumptions are not consistent with the above estimate for the seed field.
    To avoid this incongruence, 
    one could explicitly redefine the averaging procedure in the mean-field model using 
    e.g. Gaussian filtering or a form of averaging that approximates that used in observations or simulations
    \citep[e.g.][]{Zhou+18}.
    
    \subsection{Small-Scale Magnetic Fields}
    \label{sec:ss}
    
    The fluctuating or small-scale component of the field 
    is believed to be amplified by a fluctuation or small-scale dynamo,
    which is more ubiquitous than the mean-field dynamo in that it 
    does not require large-scale stratification, rotation or shear,
    nor mean small-scale kinetic helicity
    \citep{Biermann+Schluter51,Kraichnan+Nagarajan67,Kazantsev68,Meneguzzi+81,Kulsrud+Anderson92,Brandenburg+Subramanian05a,Brandenburg+12a}.
    The exponentiation time of the small-scale field in the linear regime of the fluctuation dynamo
    is  much smaller than that of the mean-field dynamo,
    and the small-scale field would have already saturated while the large-scale field is still in the kinematic regime 
    \citep{Kapyla+08,Pietarilagraham+12,Subramanian+Brandenburg14,Bhat+19}.
    Hence, a reasonable approach for modeling the small-scale  magnetic field
    is to assume a value for $b/B\eq=(\mean{\bfb^2})^{1/2}/B\eq$ that is consistent 
    with the saturated state in small-scale dynamo DNSs 
    \citep[e.g.][]{Federrath+11,Kim+Ostriker15b} and analytic models \citep[e.g.][]{Schober+15}.
    Some dependence of $b/B\eq$ on ISM parameters like the Mach number can also be extracted from those models.
    However, $b/B\eq$ would be different in the presence of a near-equipartition (with turbulent kinetic energy) large-scale magnetic field 
    generated by a large-scale dynamo (see also Sec.~\ref{sec:strength_B_tot}).
     
    The small-scale or fluctuating component of the magnetic field is also relevant for the large-scale or mean-field dynamo.
    We already mentioned (Sections \ref{sec:galactic_dynamo_theory} and \ref{sec:strength_theory}) 
    the role of the mean small-scale magnetic helicity, 
    which grows to become important as the mean field approaches $B\eq$.
    We also mentioned how the small-scale field can seed the large-scale field. 
    But the small-scale field amplified by the fluctuation dynamo or injected by SNe may 
    affect the large-scale field evolution in other ways, as well.
    
    In their mean-field dynamo model, 
    Moss et~al.~\citep{Moss+12,Moss+13,Moss+15} injected mean magnetic field with spatial scale $\sim100\pc$  
    varying randomly every $\sim10\Myr$ into the flow into regions covering $\sim1\%$ of the disc, 
    to mimic the generation of small-scale field by SNe and small-scale dynamo action. 
    Since the injected field is a perturbed part of the mean field,
    fluctuating and mean field components are entangled, 
    in contrast to their separation in standard theory (Sec.~\ref{sec:theory_def}).
    
    A different approach is to include extra terms in the mean electromotive force (emf), derived from first principles,
    that depend on the strength or other properties of the small-scale magnetic field.
    Such a galactic dynamo model was developed in Chamandy and Singh \citep{Chamandy+Singh17,Chamandy+Singh18},
    using a mean emf generalized to include the effects of rotation and stratification
     \citep{Radler+03,Brandenburg+Subramanian05a}    
    and  feedback  that accounts for turbulent tangling \citep{Rogachevskii+Kleeorin07} as a source of $\bfb$. 
    This leads to a new  quenching that is competitive with dynamical $\alpha$-quenching for typical galaxy parameter values.
    However, the theory on which the model is based still needs to be generalized, 
    for example to include the effect of shear on the mean emf, and tested using DNSs.
    
    Since there is ultimately one induction equation for the total magnetic field, 
    the mean field and fluctuation dynamos operate contemporaneously and are coupled.  
    Aspects of their separability and inseparability continue to be studied \citep{Subramanian99,Subramanian+Brandenburg14,Bhat+16,Bhat+19}.
    
    Anisotropic turbulent fields on the energy dominating scales of turbulence
    may arise from otherwise isotropic forcing 
    by e.g. a background density gradient or ordered shear, e.g. from global differential rotation. 
    A non-linear MHD cascade  also produces anisotropic turbulence 
    on progressively smaller and smaller scales \cite{Shebalin+1983,Goldreich+1994} with respect to the dominant local magnetic field,  
    but this anisotropy is likely subdominant for energy dominating scales. 
    Moreover,  the direction of the energy containing eddy scale field is itself largely random, which washes out anisotropy as measured in the observer frame.
    
    \citet{Blackman98} offered a conceptual alternative to traditional mean field dynamos 
    by instead appealing to the rms average of exponentially amplified small but finite scale fields from the fluctuation dynamo 
    then sheared in the global flow to achieve synchrotron polarization.
    This does not rely on the presence of an $\alpha$ effect 
    because shearing of the small-scale injected field alone provides a large-scale (i.e. regular) field. 
    This is a viable mechanism to produce the anisotropic turbulent field, but may predict too weak a regular field with too many reversals to  universally account for the observed regular fields of galaxies.
    
    We can estimate the regular field strength that might be expected from this sheared turbulent field alone.
    We write the turbulent correlation length of the small-scale magnetic field as $l_b$,
    and estimate the corresponding volume as $v\sim\tfrac{4}{3}\uppi l_b^3(1+\Omega\tau)$, 
    where the latter expression accounts for the stretching along $\bm{\hat{\phi}}$.
    Further, we can write $l_b=f_b l$,
    where $l$ is the correlation length of the fluctuating component of the velocity field.
    Studies involving fluctuation dynamo DNSs performed on a Cartesian mesh typically obtain values for $f_b$ in the range $1/3$ to $1/2$ in the saturated state \citep{Bhat+Subramanian13,Sur+18}.
    The galaxy has volume $V\sim 2\uppi R^2 h$. 
    The strength of the global $m=0$ component of the regular magnetic field $B\f$ averaged over the galaxy can be estimated as $\sim b/(V/v)^{1/2}$.
    Adopting typical values $l=100\pc$, $\Omega\tau=0.3$, $R=15\kpc$ and $h=400\pc$ gives $B\f\approx 3\times10^{-3}f_b^{3/2}\,b$$\approx(0.6$--$1)\times10^{-3}b$.
    If $B\f$ is obtained by averaging only over the annulus $r=2.4$--$3.6\kpc$, then we instead obtain $B\f\approx0.02f_b^{3/2}\,b$$\approx(4$--$8)\times10^{-3}b$.
    Finally, if we repeat the last estimate but use slightly different but still plausible parameter values $h=200\pc$ (reasonable for a flared disc),
    $l=200\pc$ (reasonable if superbubbles are important in driving turbulence), and $\Omega\tau=1$, we obtain $B\f\approx0.09 \,f_b^{3/2}\approx(0.02$--$0.03)b$.
    These numbers can be compared with those in Table~\ref{tab:strength}.
    We see that these estimates are probably too low to account for values of $B_\mathrm{reg}/B_\mathrm{tot}$ 
    in galaxies like M\,31, M\,33, and M\,51, but such a model might work in some galaxies.
    Note that the above estimates invoke a 3-D volume average.
    If we were to use a 1-D line-of sight average,
    then the values could be higher.
    This again highlights the need for theorists and observers to agree on the averaging method.
    
    Assessing whether galaxies that display a finite but weak regular magnetic field have a subcritical, and thus inactive, large-scale $\alpha\Omega$ dynamo warrants further work.
    
    \section{Magnetic Pitch Angle}
    \label{sec:pitch}
    
    \subsection{Observations}
    \label{sec:pitch_observations}
    
    The common definition of the pitch angle includes a sign that indicates trailing (-) or leading (+) spirals.
    As all 19 galaxies investigated in this work host trailing spiral arms and magnetic field lines, 
    we simply neglect the sign in the following.
    The pitch angles $p_\mathrm{s}$ of the spiral arms, observed in optical light, dust, or gas, 
    vary strongly between individual arms and also with distance from a galaxy's centre. 
    Table~\ref{tab:pitch} gives estimates from various tracers and the corresponding references, e.g. from a Fourier analysis of the structure of HII regions \cite{Puerari+92}.
    
    The pitch angle $p_\mathrm{o}$ of the ordered magnetic field in the galaxy's plane is computed from the observed polarization angle in the sky plane, corrected for Faraday rotation (see e.g. Fig.~16 in \citet{Beck07}). If a proper correction of Faraday rotation is not possible (because no multi-frequency data sets are available), then the apparent polarization angle, averaged over all azimuthal angles, can still be used because the average Faraday rotation measure is small for any large-scale magnetic mode. A high frequency should be used to reduce the effect of Faraday rotation. Using the compilation by \citet{Oppermann+15}, we estimate that the average Faraday rotation in the foreground of the Milky Way on the angular scales of nearby galaxies causes a constant offset in the apparent pitch angle $p_\mathrm{o}$ of less than $10^\circ$ if the angular distance between the galaxy and the Galactic plane is larger than about $10^\circ$.
    
\begin{figure*}[t]
    \vspace*{7mm}
    \begin{center}
    \includegraphics[width=13cm]{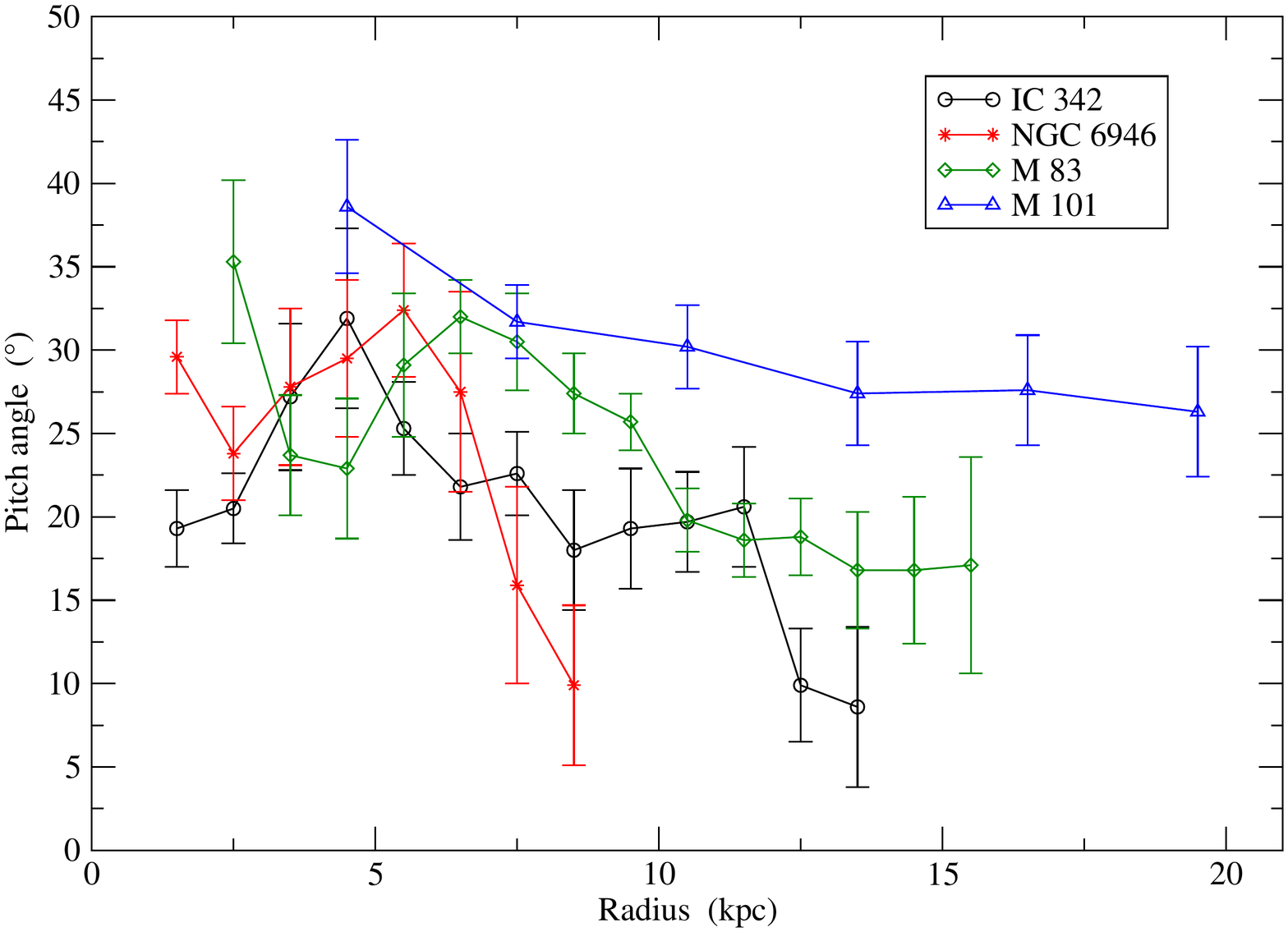}
    \caption{Radial variation of the pitch angle $p_\mathrm{o}$ (absolute values) of the ordered field, computed from the original maps in Stokes $Q$ and $U$, averaged over all azimuthal angles of each radial ring in the galaxy's plane. For NGC~6946 \citep{Beck07} data at 8.46\,GHz and 4.86\,GHz (3.6\,cm and 6.2\,cm wavelengths) allowed us to correct the pitch angles for Faraday rotation, while those for IC~\,342 \citep{Beck05}, M\,83 \citep{Frick+16}, and M\,101 \citep{Berkhuijsen+16} are based on apparent polarization angles at 4.86\,GHz. The low angular resolution of the M\,101 observations corresponds to a spatial resolution of about 5\,kpc at the adopted distance, so that the radial variation is smeared out. The spatial resolution for the other three galaxies is 0.4--0.6\,kpc.}
    \label{fig:pitch}
    \end{center}
    \end{figure*}

    Table~\ref{tab:pitch} lists the average pitch angles $p_\mathrm{o}$ of the ordered field in 19 spiral galaxies for which sufficient data are available. In the four best-studied spiral galaxies, $p_\mathrm{o}$ remains roughly constant with values between about $20^\circ$ and $35^\circ$
    until several kpc in radius, i.e. in the region of strong star formation, followed by a decrease at larger radii (Fig.~\ref{fig:pitch}).
    M\,51 is not included in this figure because the pitch angles at large radii reveal strong azimuthal variations due to the presence of the companion galaxy.
    
    \begin{table}
        \caption{Pitch angles of the spiral arms $p_\mathrm{s}$ (observed in optical, CO, or HI emission, see references), of the ordered field $p_\mathrm{o}$, and of the regular (mean) field $p_\mathrm{B}$, averaged over all azimuthal angles in the quoted radial range. The error of $p_\mathrm{o}$ is the error of the mean value over all azimuthal angles; the dispersion of $p_\mathrm{o}$ is about $4\times$ larger. Measurements of $p_\mathrm{B}$ are based on mode fitting of polarization angles (M) or on fitting the azimuthal $RM$ variation (RM). If several modes were found, the pitch angle of the axisymmetric ($m=0$) mode is given. This is an updated and extended version of Table~2 in \citet{Vaneck+15}.}
    \label{tab:pitch}
    \vspace{0.2cm}
    \centering
    \begin{tabular}{l l l l l l l l}
    \hline
    Galaxy & Distance & $p_\mathrm{s}$ [$^\circ$] & Radial range &	$p_\mathrm{o}$ $^a$ & $p_\mathrm{B}$ & Method & Reference\\
           & [Mpc] & and ref. & [kpc]   & [$^\circ$] & [$^\circ$] &   & \\
    \hline
    M\,31  & 0.78 & 7--8 & 6.8--9.0 $^b$   & --  & $13\pm4$  & M & \cite{Fletcher+04}\\
    (NGC\,224) & & \cite{Nieten+06} & 9.0--11.3 $^b$  & --  & $19\pm3$  & M \\
           & & & 11.3--13.6 $^b$ & --  & $11\pm3$  & M \\
           & & & 13.6--15.8 $^b$ & --  & $8\pm3$  & M \\
           & & & 7.0--8.0   & $30\pm5$ & $4\pm5$  & RM & \cite{Beck+19}\\
           & & & 8.0--9.0   & $29\pm4$ & $9\pm3$  & RM \\
           & & & 9.0--10.0  & $26\pm3$ & $7\pm3$  & RM \\
           & & & 10.0--11.0 & $27\pm2$ & $7\pm2$  & RM \\
           & & & 11.0--12.0 & $27\pm3$ & $5\pm3$  & RM \\
    M\,33  & 0.84 & 29--50 & 1.0--3.0   & $48\pm5$ $^{c}$ & $51\pm2$ $^d$ & M & \cite{Tabatabaei+08}\\
    (NGC\,598) & & \cite{Sandage+80} & 3.0--5.0   & $40\pm5$ $^{c}$ & $41\pm2$ $^d$ & M \\
           & & & 5.0--7.0   & $41\pm5$ $^{c}$ & -- & -- \\
           & & & 7.0--9.0   & $35\pm6$ $^{c}$ & -- & -- \\
    M\,51  & 7.6 & 15--25 & 1.2--2.4   & $20\pm2$        &  -- & -- & \cite{Fletcher+11}\\
    (NGC\,5194) & & \cite{Patrikeev+06} & 2.4--3.6   & $27\pm2$        & $20\pm1$ $^d$ & M \\
           & & & 3.6--4.8   & -- ~$^{e}$ & $24\pm4$ $^d$ & M \\
           & & & 4.8--6.0   & -- & $22\pm4$ $^d$ & M \\
           & & & 6.0--7.2   & -- & $18\pm1$ $^d$ & M \\
           & & & 7.2--8.4   & $19\pm5$ $^{e}$   & -- & -- &\\
    M\,74  & 7.3 & 13--41 & 4.0--5.0 west & $61\pm4$ $^c$ & --  &  -- & \cite{Mulcahy+17}\\
    (NGC\,628) & & \cite{Puerari+92} & 8.0--9.0 west & $45\pm2$ $^c$ & --  &  --\\
           & & & 4.0--5.0 east & $19\pm3$ $^c$ & --  &  --\\
           & & & 8.0--9.0 east & $24\pm4$ $^c$ & --  &  --\\
    M\,81  & 3.25 & 14--24 & 6.0--9.0   & $21\pm7$ & $6\pm6$  & RM & \cite{Krause+89b}\\
    (NGC\,3031) & & \cite{Puerari+92} & 9.0--12.0  & $26\pm6$ & $20\pm4$ & RM \\
           & & & 6.0--9.0   & --       & $14\pm13$ & M & \cite{Sokoloff+92}\\
           & & & 9.0--12.0  & --       & $14\pm20$ & M \\
    M\,83  & 8.9 & 14--17, $\approx$10 & 2.0--3.0   & $35\pm5$ $^c$ & --  &  -- & Fig.~\ref{fig:pitch}\\
    (NGC\,5236) & & \cite{Frick+16,Puerari+92} & 4.0--5.0   & $23\pm4$ $^c$ & --  &  -- \\
           & & & 6.0--7.0   & $32\pm2$ $^c$ & --  &  -- \\
           & & & 10.0--11.0 & $20\pm2$ $^c$ & --  &  -- \\
    M\,101 & 7.4 & 10--30 & 3.0--6.0   & $39\pm4$ $^c$ & --  &  -- & Fig.~\ref{fig:pitch}\\
    (NGC\,5457) & & \cite{Berkhuijsen+16} & 9.0--12.0  & $30\pm3$ $^c$ & --  &  -- \\
           & & & 15.0--18.0 & $28\pm3$ $^c$ & --  &  -- \\
    NGC\,253  &3.94 & ? $^f$ & $\approx$2--12 & $25\pm5$  & $26\pm7$ $^e$ & RM  & \cite{Heesen+09b}\\
    NGC\,1097 & 17.0 & 27--35 & 1.25--2.5 & -- ~$^{g}$   & $34\pm7$  & M   & \cite{Beck+05}\\
              & & \cite{Yu+Ho18} & 2.5--3.75 & --          & $36\pm5$  & M   &\\
              & & & 3.75--5.0 & --          & $23\pm2$  & M   &\\
    NGC\,1365 & 18.6 & ? $^f$ & 2.625--4.375 & -- ~$^{g}$& $34\pm2$  & M   & \cite{Beck+05}\\
              & & & 4.375--6.125 & --       & $17\pm1$  & M   &\\
              & & & 6.125--7.875 & --       & $31\pm1$  & M   &\\
              & & & 7.875--9.625 & --       & $22\pm1$  & M   &\\
              & & & 9.625--11.375 & --      & $37\pm4$  & M   &\\
              & & & 11.375--13.125 & --     & $29\pm11$ & M   &\\
              & & & 13.125--14.875 & --     & $33\pm6$  & M   &\\
    NGC\,1566 & 17.4 & 19--21 & 2.0--4.0 & $29\pm4$ $^c$ & -- & -- & \cite{Ehle+96}\\
              & & \cite{Yu+Ho18} & 4.0--6.0 & $17\pm4$ $^c$ & -- & -- \\
              & &                   & 6.0--8.0 & $15\pm4$ $^c$ & -- & -- \\
    NGC\,3627 & 11.9 & 10--50 & $\approx$2--5 west& $16\pm2$      & -- & -- & \cite{Soida+01}\\
              & & \cite{Soida+01} & $\approx$2--5 east& $27\pm2$      & -- & -- &\\
              & & & 4.0--7.0 east& $68\pm4$ $^{g,h}$  & -- & -- &\\
      \hline
    \end{tabular}
    \end{table}
              
    \begin{table}
        \caption{Table \ref{tab:pitch} continued}
    \vspace{0.2cm}
    \centering
    \begin{tabular}{l l l l l l l l}
    \hline
    Galaxy & Distance &  $p_\mathrm{s}$ [$^\circ$] & Radial range & $p_\mathrm{o}$ & $p_\mathrm{B}$ & Method & Reference\\
           & [Mpc] & and ref. & [kpc]   & [$^\circ$] & [$^\circ$] &   & \\
    \hline
    NGC\,4254 & 16.8 & 6--72 & 1.2--2.4     &  $28\pm4$  & --  & -- \\
              & & \cite{Iye+82} & 2.4--3.6     &  $27\pm3$  & --  & -- \\
              & & & 3.6--4.8     &  $30\pm3$  & --  & -- \\
              & & & 4.8--6.0     &  $26\pm3$  & $27\pm9$ $^e$   & RM & \cite{Chyzy08}\\
              & & & 6.0--7.2     &  $22\pm3$  & $24\pm5$ $^e$  & RM \\
              & & & 7.2--8.4     &  $24\pm3$  & $29\pm7$ $^e$  & RM \\
    NGC\,4414 & 19.2 & 20--40 & $\approx$4--7& -- ~$^g$ & $\approx$22  & M & \cite{Soida+02}\\
              & & \cite{Thornley+Mundy97} & 1.5--3.0     & $30\pm5$   & --  &  -- \\
              & & & 3.0--4.5     & $27\pm4$   & --  &  -- \\
              & & & 4.5--6.0     & $27\pm4$   & --  &  -- \\
              & & & 6.0--7.5     & $26\pm5$   & --  &  -- \\
    NGC\,4449 & 3.7 & -- $^i$ & 1.0--2.0 & -- ~$^g$    & $59\pm11$ $^e$  & RM & \cite{Chyzy+00}\\
              & & & 2.0--3.0     & $28\pm7$   & $38\pm9$ $^e$ & RM \\
    NGC\,4736 & 4.66 & ? $^f$ & $\approx$0.3--3.0 & $35\pm5$   & --  & --  & \cite{Chyzy+Buta08}\\
    NGC\,6946 & 7.0 & 20--28 & 0.0--6.0 & $27\pm2$   & --  & --  & \cite{Ehle+Beck93}\\
              & & \cite{Puerari+92} & 6.0--12.0    & $21\pm2$   & --  & -- \\
              & & & 12.0--18.0   & $10\pm6$   & --  & -- \\
              & & & 1.0--2.0      & $30\pm2$ $^c$ & --  & -- & Fig.~\ref{fig:pitch}\\
              & & & 5.0--6.0      & $32\pm4$ $^c$ & --  & -- \\
              & & & 8.0--9.0      & $10\pm5$ $^c$ & --  & -- \\
    IC\,342   & 3.4 & 10--25 & 5.5--9.9 $^j$ & $22\pm2$   & $\approx0$ & RM  &\cite{Grave+Beck88}\\
              & & \cite{Crosthwaite+00} & 9.9-14.3 $^j$ & $16\pm2$  & $6\pm10$  & RM \\
              & & & 5.5--9.9  $^j$ & $20\pm2$   & $34\pm18$ & RM & \cite{Krause+89a}\\
              & & & 9.9--14.3 $^j$ & $16\pm2$   & $7\pm9$   & RM \\
                            & & & 5.5--9.9  $^j$ & --     & $20\pm4$  & M & \cite{Sokoloff+92}\\
              & & & 9.9--14.3 $^j$ & --     & $16\pm11$  & M \\
              & & & 7.5--12.5   & $19\pm2$ $^c$ & $30\pm10$  & RM & \cite{Beck15a}\\
              & & & 12.5--17.5  & $10\pm2$ $^c$ & $4\pm14$  & RM \\
              & & & 1.0--2.0    & $19\pm2$ $^c$ & --  & -- & Fig.~\ref{fig:pitch}\\
              & & & 5.0--6.0    & $25\pm3$ $^c$ & --  & -- \\
              & & & 8.0--9.0    & $18\pm4$ $^c$ & --  & -- \\
              & & & 12.0--13.0  & $10\pm3$ $^c$ & --  & -- \\
    LMC       & 0.05 & ? $^f$ & 0--3 & --           & $4\pm10$ & \textit{RM} $^k$  & \cite{Mao+12b}\\
    \hline
    \end{tabular}
    \vspace{0.2cm}
    \footnotesize
    \begin{itemize}
    \item[]
    $^a$ Re-computed from the original maps in Stokes $Q$ and $U$ at two frequencies (if available), including\\
    correction for Faraday rotation; 
    $^b$ scaled to 780\,kpc distance;
    $^c$ re-computed from the original maps in\\
    Stokes $Q$ and $U$ at one frequency;
    $^d$ using the $m=0$ result of the fit for modes 0+2;\\
    $^e$ re-analyzed from the original data;
    $^f$ no data available;
    $^g$ strong variations with azimuthal angle;\\
    $^h$ magnetic arm with anomalously large pitch angle;
    $^i$ irregular dwarf galaxy;
    $^j$ scaled to 3.4\,Mpc distance;\\
    $^k$ from $RMs$ of background sources.
    \end{itemize}
    \normalsize
    \end{table}

    As mentioned in Section~\ref{sec:intro}, the observed ordered field may include anisotropic turbulent fields and hence $p_\mathrm{o}$ may be different from the pitch angle $p_\mathrm{B}$ of the regular field. Turbulent fields are strongest in the inner disc, where star formation is strongest, while regular fields may extend to much larger radii. Hence, it cannot be excluded that the radial decrease of the pitch angle in Fig.~\ref{fig:pitch} is due to anisotropic turbulent fields.
    
    Measuring the average pitch angle $p_\mathrm{B}$ of the regular field needs a mode analysis of the polarization angles at several frequencies \citep[e.g.][]{Fletcher+11}. Results obtained from this method are rare (see Sec.~\ref{sec:non-axisymmetric}). 
    Alternatively, one may use the phase shift of the sinusoidal fit to the azimuthal variation of $RM$ in the case that the axisymmetric mode dominates. 
    Table~\ref{tab:pitch} summarizes the presently available data.
    
    A significant radial decrease of $p_\mathrm{B}$ was found only in M\,31 and M\,33.
    However, the most recent $p_\mathrm{B}$ values for M\,31 derived with the $RM$ method are smaller than the previous ones derived with the mode analysis method (lines $1-4$ in Table~\ref{tab:pitch})  
    and do not show a significant radial variation. The reason could be that either the $RM$ method is unreliable due to the presence of higher modes or that the mode analysis method is unreliable due to strong anisotropic turbulent fields. This question needs to be investigated in more detail.
    
    A correlation between the pitch angles of the spiral magnetic patterns and the spiral structure of the optical arms $p_\mathrm{s}$ would suggest that the processes of the formation of both phenomena are related or that these processes interact with each other.
    Such a correlation has been suggested but not yet firmly established \citep{Vaneck+15}.
    A similarity in the magnitudes of the two types of pitch angle would also suggest a physical connection.
    
    In the galaxies for which data of $p_\mathrm{o}$ and $p_\mathrm{B}$ are available, $p_\mathrm{o} \simeq p_\mathrm{B}$ or $p_\mathrm{o} > p_\mathrm{B}$ is valid for the azimuthally averaged pitch angles. 
    The largest difference is found in M\,31. 
    For individual magnetic arms, $p_\mathrm{o}$ was found to be larger than $p_\mathrm{s}$ by about $20^\circ$ in M\,83 \cite{Frick+16}, about $8^\circ$ in M\,101 \cite{Berkhuijsen+16}, and $12^\circ - 23^\circ$ for the three main arms in M\,74 \cite{Mulcahy+17}.
    This could indicate that anisotropic turbulent fields, responsible for a significant fraction of the polarized emission, have a systematically larger pitch angle than the regular field.
   
    \subsection{Magnetic Pitch Angle from Dynamo Models}
    \label{sec:pitch_theory}
    
    In mean-field galactic dynamo models, the pitch angle of the mean magnetic field $p$ depends, 
    in the saturated state, only weakly on the details of the non-linear dynamo quenching.
    For this reason, it can be written down as a simple expression that can be shown 
    to agree rather closely with numerical solutions \citep{Chamandy16,Chamandy+16}:
    \begin{equation}
      \label{pitch}
      p \simeq \arctan\left(\frac{\pi^2\,\tau\,u^2+6\,h\, U\f}{12\,q\shear\,\Omega\, h^2}\right),
    \end{equation}
    where $U\f$ is the characteristic outflow speed (see Sec.~\ref{sec:strength_theory}).
    In \citet{Chamandy+16}, magnetic pitch angles $p_\mathrm{B}$ for a handful of nearby galaxies 
    were compared with the results of a simple dynamo model, 
    which had only a single free parameter, $\tau$, with other parameters constrained by observations.
    A constant turbulent speed $u$ of $10\kms$ was assumed.
    The level of agreement was reasonable 
    and required $\tau\sim14\Myr$, close to the canonical order of magnitude estimate of $10\Myr$.
    The scale height $h$ was modeled as an increasing function of $r$; 
    a model with constant $h$ provided a much poorer fit to the data,
    regardless of the value of $h$ adopted.
    This is an example of how magnetic fields can be used to indirectly \textit{probe} other galactic properties.
    Flaring of galactic discs is expected on theoretical grounds \citep{Banerjee+11,Bacchini+18}, 
    and has been observed in warm neutral (\hi) gas in the Milky Way \citep{Kalberla+Kerp09}
    and in other galaxies \citep{Yim+11,Yim+14,Peters+17}. Flaring of the disc of warm ionized gas is not evident from observations in the Milky Way \citep{Yao+17}, and no data are available for external galaxies yet.
    
    On the other hand, equation~\eqref{pitch} tells us that a radially decreasing turbulent speed $u$, 
    which is suggested by the \hi\ line profiles in many spiral galaxies \citep{Tamburro+09,Ianjamasimanana+15}
    (see also Sec.~\ref{sec:velocity_dispersion_radial})
    could affect the radial variation of $p$ in the same direction as flaring,
    and no attempt has yet been made to unravel this possible degeneracy in the models.
    Better data are needed; the updated and improved data set for the regular magnetic fields of nearby galaxies presented in this work 
    provides a first step in this direction. 
    
    That $p_\mathrm{o}$ is found to be similar to or larger than $p_\mathrm{B}$ is broadly consistent with theoretical expectations.
    The pitch angle of the anisotropic fluctuating field is expected to be of the order $\arctan[1/(1+q\shear\Omega\tau)]$,
    since shear from galactic differential rotation can affect the properties of the field  over the turbulent correlation time.
    Typically $q\shear\Omega\tau \lesssim1$.
    A similar argument for the mean field gives $\arctan[1/(1+q\shear\Omega t_\mathrm{d})$,
    where $t_\mathrm{d}=h^2/\etat\simeq 3h^2/(\tau u^2)$ is the vertical turbulent diffusion time, and
    where we have made use of equation~\eqref{etat}.
    Here $\Omega t_\mathrm{d} \gg 1$, and this estimate agrees well with equation~\eqref{pitch} as long as $U\f$ is not too large.
    This is why mean-field pitch angles are expected to be smaller than anisotropic fluctuating field pitch angles. 
    The same argument was used to make a similar prediction for accretion disks \cite{Blackman+Nauman15}.
    
    Non-axisymmetric forcing of the dynamo by a spiral modulation of a parameter like $\alpha$
    can lead to a spiral modulation of the pitch angle,
    as indicated by some observations \cite{Rohde+99},
    but the pitch angle of the azimuthally averaged mean field
    tends to be insensitive to the pitch angle of the forcing spiral \citep[e.g.][]{Chamandy+13a}.
    However, these results are from models that do not include non-axisymmetric mean velocity fields
    or possible feedback on the spiral structure from the dynamo-generated mean magnetic field,
    and thus more work is needed to establish whether a causal relationship between $p$ and the pitch angle
    of the underlying spiral structure of the galaxy $p_\mathrm{s}$ should be expected.
    Even so, any correlation between $p_\mathrm{B}$ and $p_\mathrm{s}$ need not imply a causal relationship,
    because $p_\mathrm{s}$ is known to be anti-correlated with the galactic shear rate $q\shear$ \citep{Seigar+05,Seigar+06,Grand+13},
    and an anti-correlation between $p_\mathrm{B}$ and $q\shear$ is suggested by the theory for $p$ (equation~\ref{pitch}),
    and also by the data \citep{Chamandy+16}.
    Likewise, one would also expect $p_\mathrm{o}$ to be smaller for larger $q\shear$, 
    when the former is dominated  by the anisotropic turbulent component, if this component is generated by global shear.
    In summary, from dynamo models, some correlation between the pitch angles of the field
    and spiral arms is expected, but not necessarily because of a direct interaction between the two.
    
    \section{Statistical Correlations}
    \label{sec:correlations}
    
    Other attempts have been made to statistically analyze magnetic field and other data to assess possible correlations between observables.
    
    \citet{Vaneck+15} identified nine such statistical correlations (out of 23 pairs of observables considered).
    This includes a correlation (of particularly high significance) between the amplitude and pitch angle of the regular axisymmetric mode
    for the five galaxies for which the mode analysis had been carried out, 
    with a fitted relation of the form $\tan p_\mathrm{B} \propto (-0.51\pm0.11) \times \log B_\mathrm{reg}$ 
    with a correlation coefficient of $-0.56$. 
    This is roughly consistent with our revised data for six galaxies\,
    \footnote{NGC\,253 and IC\,342 are excluded because of their exceptionally weak regular fields.}
    that yield a slope of $-1.3\pm0.4$ and a similar correlation coefficient of $-0.64$.
    This result needs to be verified with more data and explained using dynamo models.
    
    In another statistical study, Chy\.zy et al. \citep{Chyzy+11,Chyzy+17} 
    found correlations between total field strength and specific star formation rate,
    and between the former and Hubble type.
    \citet{Tabatabaei+16} argued for a correlation betweeen the strength of the ordered field and the galactic rotation speed
    (averaged over the flat part of the rotation curve).
    They derive $B_\mathrm{ord}\propto v_\mathrm{rot}^{~~0.7-1.7}$, where the uncertainty in the exponent 
    accounts for the choice of the method used to calculate the number of CR electrons in the integration volume.
    Larger statistical studies will become feasible with up-and-coming instruments, 
    and models are needed now to explain the above results and predict future results.

    \section{Halo Magnetic Fields}
    \label{sec:halo}
    
    \subsection{Observations}\label{sec:halo_obs}
    
    Edge-on views of most spiral galaxies reveal a thin and a thick
    disc in synchrotron emission with exponential scale heights of a few 100\,pc and a few kpc, respectively \citep{Heesen+18,Krause+18}.
    Thick radio discs are also called ``radio halos''.
    Very few galaxies (e.g. M~31) have only a thin disc in synchrotron emission. 
    In most galaxies, the synchrotron scale height of the halo is $H_\mathrm{syn}=1-2$\,kpc and the disc scale length is $R_\mathrm{syn}=3-6$\,kpc \citep{Krause+18}. 
   
    Emission at greater heights above the galactic midplane may exist, but its detection would need more sensitive radio observations 
    because cosmic-ray electrons propagating away from their places of origin in the disc into the halo 
    lose most of their energy within a few kpc.
    Nevertheless, the dependence of synchrotron intensity on $B^2$ and on density of cosmic-ray electrons implies that
    magnetic field scale heights are $2-4$ times (depending on synchrotron losses) as large as $H_\mathrm{syn}$,
    so $H_\mathrm{B}\simeq2$-$8\kpc$.

    \citet{Krause09} found that the ordered fields are ``X-shaped'' in most galaxy halos, i.e. they have a strong $z$ component.
    The strength of the ordered field in the halo is often similar to that of the ordered disc field near the disc plane.
    Due to the lack of good $RM$ data, we cannot say much about the $z$-component of the regular field nor about field reversals at certain heights above the disc plane.
    A few edge-on galaxies, NGC\,4631 \citep{Mora+19}, NGC\,4666 \citep{Stein+19a}, 
    and other cases in the CHANG-ES sample show strong $B_{\mathrm{reg},z}$ in the halo.
    The regular field in the halo of NGC\,4631 is coherent over about 2\,kpc in height, reverses on a scale of about 2\,kpc in radius, and is about $4\,\mu$G strong, compared to the average strength of the total field in the disc of about $9\,\mu$G.
    
    \subsection{Dynamo Models}
    \label{sec:halo_theory}
    
    Here we summarize the models for halo magnetic fields, 
    focussing on global mean-field dynamo models;  
    we do not discuss local ISM models which extend partway into the halo \citep[e.g.][]{Gent+13b},
    nor other types of global models \citep[e.g.][]{Hanasz+09b,Rieder+Teyssier17a}.
    For a more extensive review of magnetic fields around (as opposed to in) galactic discs, see \citet{Moss+Sokoloff19}.
    
    That $\!\mkG$ strength fields exist in the relatively tenuous halos of galaxies is  plausible,
    given that the ratio of equipartition field strengths (equation~\ref{Beq}) in halo and disc is
    $(B\halo/B\disk)=(\rho\halo/\rho\disk)^{1/2}(u\halo/u\disk)\approx0.4$,
    using $\rho\halo\approx10^{-2}\rho\disk$ and $u\halo\approx4u\disk$.
    The expected ratio of radio intensities is $(I\halo/I\disk)=(B\halo/B\disk)^{2...4}\approx0.02...0.2$, 
    depending on the level of correlation between the energy densities of magnetic fields and CRs, 
    whereas observations often show radio halos brighter than a ratio of 0.2.

    In existing mean-field dynamo models, discussed below,
    an unflared disc is embedded in a spherical halo.
    The dynamo parameters vary smoothly, but abruptly, between these regions.
    The turbulent correlation length $l$ and rms turbulent velocity $u$ are both larger in the halo 
    than in the disc, leading to substantially larger turbulent diffusivity 
    ($\propto lu$ assuming $\tau\sim l/u$; see equation~\ref{etat}) in the halo. 
    The $\alpha$ effect can also take on different values in disc and halo. 
    If a wind is present, it transitions from vertical in the disc to radially outward in the halo.
    
    ``X-shaped'' polarization signatures in the halo have been explained
    with mean-field models that incorporate prescribed galactic winds but
    assume that the mean-field dynamo ($\alpha$ effect) in the halo is weak or absent \citep{Brandenburg+93}.
    Then field possesses even (quadrupole-like) symmetry everywhere,
    the same symmetry that is obtained in pure disc dynamo solutions (Sec.~\ref{sec:parity}).
 
    Alternatively, the \textit{halo} may drive a mean-field dynamo 
    whose dominant eigenmodes are of odd (dipole-like) symmetry 
    and maybe oscillatory \citep{Moss+Tuominen90,Sokoloff+Shukurov90,Brandenburg+92}.
    Then, mixed parity field solutions which are even in the disc and odd in the halo are possible.
    \citet{Brandenburg+92} found that global eigensolutions are usually of purely even or odd parity,
    depending on whether the disc or halo dynamo dominates (and so enslaves the other),
    and that this carries over into the non-linear regime as well.
    However, they find that the small growth rate of the halo field
    can prevent such pure parity solutions from emerging within a galaxy lifetime. 
    This implies  (i) mixed parity transient solutions are likely important, 
    and (ii) the disc may evolve independently of the halo.
    According to Sec.~\ref{sec:parity}, there is a clear prevalence of even parity in radio halos for the galaxies in Table~\ref{tab:parity}.
    This supports dynamo solutions for which the disc dynamo dominates over the halo dynamo.
    
    More recent work  by \citet{Moss+Sokoloff08} and \citet{Moss+10}, using higher numerical resolution, 
    also showed pure parity solutions dominated by either disc or halo modes, confirming earlier results.
    However, there exists a region of parameter space for which steady or low-amplitude oscillatory solutions 
    emerge in the non-linear regime that have even parity in the disc and odd parity in the halo.
    They also find that the disc dynamo and associated even parity  dominates as the outflow speed increases.
    
    One caveat affecting all of these models is that they make use of the heuristic algebraic $\alpha$-quenching formalism,
    rather than the more modern dynamical $\alpha$-quenching formalism (Sec.~\ref{sec:galactic_dynamo_theory}).
    Since the transport of $\alpha\magn$ between disc and halo could have important implications for the dynamo,
    models which include dynamical $\alpha$-quenching are desirable.
    Such a model has been attempted, showing only very weak large-scale halo fields \citep{Prasad+Mangalam16}.
    Magnetic fields may also affect outflows \citep{Evirgen+19},
    and mean-field models that solve for the outflow self-consistently are needed.
    
    We have seen that to ``zeroth order,'' the global features of galactic magnetic fields 
    can be explained by theory, but better comparison between data and models is still needed. 
    The models discussed generally assume the field to be axisymmetric (i.e. independent of $\phi$ in cylindrical coordinates).
    We next discuss relaxing this assumption.

    \section{Non-Axisymmetric Large-Scale Fields}
    \label{sec:non-axisymmetric}
    
    In this section, we focus on observations of azimuthally varying magnetic fields in nearby galaxies
    and their interpretation using non-axisymmetric mean-field dynamo models.
    We focus on the galactic spiral structure in galactic discs,
    leaving out the effects of galactic bars on magnetic fields, and non-axisymmetric fields in galaxy halos.
    Work on the former topic can be found in a series of papers beginning with \citet{Moss+01} and \citet{Beck+02}.
    The latter topic has been studied observationally \citep{Fletcher+11} for M\,51
    and theoretically using analytical mean-field models that assume self-similarity \citep{Henriksen17b}.
    
    Measuring Faraday rotation with high resolution and at sufficiently high frequencies, to ensure that Faraday depolarization of the disc emission is small, is crucial for measuring mean (regular) fields. 
    A large-scale sinusoidal pattern of $RM$ along azimuthal angle in a galactic plane could be a signature 
    of a dominating axisymmetric regular field generated by the $\alpha\Omega$ dynamo. 
    Several modes may be superposed, so Fourier analysis of the $RM$ variation with azimuthal angle is needed.
    The resolution and sensitivity of present-day radio observations 
    is sufficient to identify $2-3$ global azimuthal modes (Table~\ref{tab:modes}). 
    The fit delivers the approximate $RM$ amplitude and pitch angle for each mode.
    \footnote{The term ``mode'' has a slightly different meaning in observation and theory.
    In theory, it usually refers to eigen modes in the kinematic regime of dynamo action.}
    
    \begin{table}
    \caption{Decomposition of regular magnetic fields in galaxy discs 
    into azimuthal modes of order $m$, derived from radio polarization observations at high frequencies, where the discs are not Faraday depolarized.
    These modes were found in several radial ranges of each galaxy.
    Columns $2-4$ give the amplitudes relative to the strongest mode, averaged over the investigated radial ranges.\\
    A ``?'' indicates modes that were not sought, whereas a zero indicates that no signature of this mode was found.
    The entries ``1 ? ?'' denote cases for which only the $m=0$ mode was fitted, but it should be noted that the reduced $\chi^2$ value of the fit was large in some cases, implying that the $m=0$ mode was not alone sufficient to explain the observed RM variation (see original references for details).
    Modes of order higher than 2 cannot be detected with the resolution of present-day telescopes. --\\
    This is a revised and extended version of a table in \citet{Fletcher10} who used results from mode fitting. Here, we also use results obtained from fitting the $RM$ variations and also more recent references.}
    \label{tab:modes}
    \begin{center}
    \begin{tabular}{l l l l l l}
    \hline
    Galaxy 	& Radial range [kpc] & $m=0$ &  $m=1$ & $m=2$ & Reference\\
    \hline
    M\,31        & 6.8--9.0 & 1 & $\approx$0   & $0.54\pm0.13$ & \cite{Fletcher+04}\\ 
                 & 9.0-15.8 & 1 & $\approx$0   & $\approx$0          & \cite{Fletcher+04}\\
    M\,31        & 9.0--11.0& 1 $^a$  & $0.18\pm0.06$ $^a$& ? & \cite{Beck+19}\\
    M\,33        & 1--5 & 1 & $\approx$0   & $0.62\pm0.07$  & \cite{Tabatabaei+08}\\
    M\,51 (disc) & 2.4--3.6 & 1 & $\approx$0   & $0.72\pm0.06$ & \cite{Fletcher+11}\\
                 & 3.6--7.2 & 1 & $\approx$0   & $0.52\pm0.07$ & \cite{Fletcher+11}\\
    M\,51 (halo) & 2.4--3.6 & $0.30\pm0.09$ & 1   & $\approx$0   & \cite{Fletcher+11}\\
                 & 3.6--7.2 & $\approx$0       & 1   & $\approx$0   & \cite{Fletcher+11}\\
    M\,81        & 9--12 & $<0.5$  & 1   & ?  & \cite{Sokoloff+92}\\
    M\,83        & 4--12 & $0.4\pm0.3$ $^a$  & 1 $^a$ & ?  & Table~\ref{tab:m83}\\
    NGC\,253     & 1.4--6.7 & 1 & ?  & ?  & \cite{Heesen+09b}\\
    NGC\,1097    & 3.75--5.0 & 1 & $0.33\pm0.05$ & $0.48\pm0.05$   & \cite{Beck+05}\\
    NGC\,1365    & 2.625--14.875 & 1 & $0.9\pm0.6$ & $0.9\pm0.3$   & \cite{Beck+05}\\
    NGC\,4254    & 4.8--7.2 & 1 & $0.58\pm0.07$ & ?  & \cite{Chyzy08}\\
    NGC\,4414    & $\approx$2-7 & 1 & $\approx$0.6 & $\approx$0.4 & \cite{Soida+02}\\
    NGC\,4449    & 1--3  & 1 & ? & ? & \cite{Chyzy+00} (re-analyzed)\\
    NGC\,6946    & 0--18 & $\approx$1 $^a$ & ? & $\approx$1 $^a$ & \cite{Ehle+Beck93,Rohde+99}\\
    IC\,342      & 5.5--17.5 & 1 & ?  & ?  & \cite{Grave+Beck88,Krause+89a,Sokoloff+92,Beck15a}\\
    LMC          & 0--3 & 1 $^a$ & ?  & ?  & \cite{Mao+12b}\\
    \hline
    \end{tabular}
    \vspace{0.2cm}
    \footnotesize
    \begin{itemize}
    \item[]
    \begin{center}
    $^a$ derived from the azimuthal $RM$ variation, not by mode fitting
    \end{center}
    \end{itemize}
    \normalsize
    \end{center}
    \end{table}

    The results of Table~\ref{tab:modes} are mostly based on data at 8.46\,GHz and 4.86\,GHz where Faraday depolarization was shown to be small \citep[e.g.][]{Beck07,Beck05,Fletcher+11}. Data at lower frequencies are affected by Faraday depolarization but can be used to investigate the field patterns in the disc and halo separately \citep{Fletcher+11}.
    
    All nearby spiral galaxies, for which sufficiently sensitive $RM$ data are available, and the dwarf galaxies NGC\,4449 and LMC reveal global large-scale $RM$ patterns (Table~\ref{tab:modes}).
    The Andromeda galaxy M\,31 is the prototype for a dynamo-generated axisymmetric spiral disc
    field, with a star-forming ring that shows one $RM$ maximum and one $RM$ minimum along azimuth, 
    strongly suggesting an axisymmetric regular spiral field (mode $m = 0$),
    with a superimposed bisymmetric spiral field (mode $m = 1$) of about $6\times$ lower amplitude \citep{Beck+19}.
    Other candidates for a dominating axisymmetric disc field are the nearby spirals IC\,342 and NGC\,253. 
    The axisymmetric regular field in the irregular Large Magellanic Cloud (LMC) is almost azimuthal (i.e. small pitch angles). 
    Dominating bisymmetric spiral fields are rare, but possibly exist in M\,81 and M\,83.
    
    The two main magnetic arms and the pattern of Faraday rotation measures in NGC\,6946 
    can be described by a superposition of two modes ($m = 0$ and $m = 2$) with about equal amplitudes, 
    where the quadrisymmetric ($m = 2$) mode is phase shifted with respect to the optical spiral arms.
    The other magnetic arms of NGC\,6946 may need additional higher modes.
    For several other galaxies, three modes ($m = 0$, 1, and 2) were found to be necessary, but not necessarily sufficient, to describe the data.
    For all galaxies for which both $m = 1$ and $m = 2$ modes were sought, the amplitude of $m = 2$ was either about as strong as that of $m = 1$ or stronger than that of $m = 1$.
    
    Mean-field dynamo models with a purely axisymmetric underlying turbulent disc can be constructed to have non-axisymmetric modes with positive growth rates.
    However, the extra spatial variation of $\meanv{B}$ of these higher modes 
    causes enhanced turbulent diffusion and smaller growth rates than the $m=0$ mode \citep{Ruzmaikin+88},
    so the latter dominates in the kinematic regime.
    If the amplitude of the $m=1$ component of the seed field were dominant, 
    the $m=1$ could saturate before $m=0$ in spite of its slower growth.
    But $m=0$ would come to dominate in the non-linear regime \citep{Chamandy+13a,Sokoloff+Moss13},
    so that non-axisymmetric near-equipartition large-scale fields likely require non-axisymmetric spiral \textit{forcing}. 
    Thus, we focus on these models below.
    
    \subsection{Magnetic Spiral Arms}
    \label{sec:magnetic_arms}
    
    The original definition of ``magnetic arms'' came from the observations of the spiral galaxy NGC\,6946 at wavelengths of 6.2\,cm and 3.6\,cm. 
    In this galaxy, the spiral arms of polarized radio emission are most prominent, located between the optical spiral arms, 
    and quite narrow, not filling the whole inter-arm space \citep{Beck+Hoernes96,Beck07}.
    The magnetic arms reveal similar patterns at both wavelengths.
    Faraday depolarization in the turbulent medium of the optical spiral arms was invoked to explain the lack of polarized emission in the optical arms. However, Faraday depolarization strongly decreases with decreasing wavelength, so that the polarized emission should become smooth and hence the magnetic arms should disappear at 3.6\,cm, which is not observed.
    
    Wavelength-independent beam depolarization occurs due to unresolved twisting/ tangling/ variation in the magnetic field.
    It is plausible that this effect could be stronger in the optical spiral arms.
    However, the total emission is also slightly enhanced in the magnetic arms, which cannot be explained by any type of depolarization. 
    $RM$ in the magnetic arms in NGC\,6946 shows that the magnetic field is mostly regular, 
    while in most other galaxies listed in Table~\ref{tab:magneticarms}, the $RM$ data are still inconclusive. 
    In conclusion, the magnetic arms are regions of enhanced ordered (probably regular) field strength that also enhances the total field strength.
    
    Investigations of many other galaxies revealed a  wide range of widths and locations relative to the optical arms.
    Hence, we extend the term ``magnetic arms'' to include all large-scale spiral structures of polarized radio emission 
    with structure pitch angles of the same sign as the optical spiral arms, but not necessarily of similar magnitudes 
    (e.g. the ``anomalous magnetic arm'' in NGC\,3627, see Table~\ref{tab:magneticarms}).
    The present status of observational results is summarized in Table~\ref{tab:magneticarms}.
    All features with roughly spiral shape and at least a few kpc lengths are counted in Table~\ref{tab:magneticarms}. 
    The lengths (measured along the arms) and the widths (i.e. full width to half intensity) are estimated from the images of polarized emission. 
    The length estimates are limited by the sensitivity of the observations and hence are lower limits. 
    Similarly, optical spiral arms are counted if at least a few kpc long.
    
    Generally speaking, grand-design and multiple arm galaxies 
    \citep[see][for definitions and classifications]{Elmegreen81,Kendall+11,Elmegreen+11} 
    tend to host magnetic arms, while irregular and dwarf galaxies (not listed in the table) do not.
    Lengths and widths of magnetic arms vary considerably among galaxies. 
    Narrow magnetic arms (about 0.5\,kpc width) exist only in optical inter-arm regions. 
    The longest magnetic arms are observed at smaller frequencies (1--3\,GHz), where the signal-to-noise ratios are higher in the outer discs, 
    while the polarized emission in the inner disc is reduced by Faraday depolarization. 
    Most magnetic arms show clear offsets from the optical spiral arms, some are coinciding with an optical arm, 
    and a few are located at the inner edges of massive optical arms, like in M\,51.
    
    \begin{table}
    \caption{Properties of magnetic arms of spiral shape in star-forming galaxies. The length is measured along the arm until the detection limit. The width is measured at half intensity. ``n'' offset means that the magnetic arms are roughly coincident with optical arms, while ``y'' indicates significant offsets. 
    }
    \label{tab:magneticarms}
    \vspace{0.2cm}
    \centering
    \begin{tabular}{l l l l l l l l l}
    \hline
    Galaxy & Class    &No.opt. &No.magn. & Length & Width & Offset to & Freq. & Ref.\\
           & (NED)   &arms    &arms     & [kpc]  & [kpc] & opt.arms  & [GHz]\\
    \hline
    M\,33     & SA(s)cd  &$\approx6$ $^a$ & 3  & 3--7 & 1     &  y/n     & 8.35  &\cite{Tabatabaei+07}\\
    M\,51     & SA(s)bc  & 2     & 4      & 7--15  & 0.5--1   &  y $^b$  & 4.86, 8.46&\cite{Fletcher+11}\\
    M\,51     &          & 2     & 4      & 10--20 & 1        &  y       & 1.5   &\cite{Mao+15}\\
    M\,74     & SA(s)c   &$\approx5$ & 5     & 6--30  & 0.5--3&  y/n     & 3.1   &\cite{Mulcahy+17}\\
    M\,81     & SA(s)ab  & 2     & 2      & 6--8   & 1.5      &  y $^b$  & 1.4   &\cite{Krause+89b}\\
    M\,83     & SAB(s)c  &$\approx5$ & 6     & 12--25 & 1--2.5&  y/n     & 2.37, 4.86&\cite{Frick+16}\\
    M\,101    & SAB(rs)cd &$\approx6$ & 2     & 30--40 & $<5$  &  y/n     & 2.70, 4.85&\cite{Berkhuijsen+16}\\
    NGC\,1097 & SB(s)b   & 2     & 2      & 4--9   & 1.5--2.5 &  y       & 4.86  &\cite{Beck+05}\\
    NGC\,1365 & SB(s)b   & 2     & 2      & 9--17  & 2--4     &  y       & 4.86  &\cite{Beck+05}\\
    NGC\,1566 & SAB(s)bc &$\approx5$ & 2  & 5--7   & $<1.5$   &  y       & 4.80  &\cite{Ehle+96}\\
    NGC\,2997 & SAB(rs)c & 3     & 4     & 7--20  & 0.5--1    &  y/n     & 4.86  &\cite{Han+99}\\
    NGC\,3627 & SAB(s)b  & 2     & 4     & 5--10  & 1.5--2    &  y/n$^c$ & 8.46  &\cite{Soida+01}\\
    NGC\,4254 & SA(s)c   & 3     & 2     & 13--15 & 1--1.5    &  y/n     & 4.86, 8.46&\cite{Chyzy+07}\\
    NGC\,4414 & SA(rs)c  &$\approx6$ $^a$& 4 & 3--10 & 1--1.5 &   y      & 8.44  &\cite{Soida+02}\\
    NGC\,4736 & (R)SA(r)ab & 0     & 0 $^d$& --     & --    & --       & 4.86, 8.46&\cite{Chyzy+Buta08}\\
    NGC\,6946 & SAB(rs)cd &5      & 4     & 6--12  & 0.5--1    &  y     & 4.86, 8.46&\cite{Beck07}\\
    NGC\,6946 &           &5      & 5     & 7--14  & 0.5--2    &  y      & 1.46  &\cite{Beck07}\\
    IC\,342   & SAB(rs)cd &$\approx5$ $^a$& 4 & 6--15  & 0.5--1&  y/n    & 4.86  &\cite{Krause93,Beck15a}\\
    IC\,342   &           &$\approx5$ $^a$& 6 & 8--30  & 1--3  &  y/n    & 1.49  &\cite{Krause+89a,Beck15a}\\
    \hline
    \end{tabular}
    \footnotesize
    \begin{itemize}
    \item[]
     $^a$ partly rudimentary arms,
     $^b$ at inner edge of optical arms,
     $^c$ one anomalous magnetic arm with a large pitch angle,\\
     $^d$ diffuse spiral pattern with large pitch angle. 
    \end{itemize}
    \normalsize
    \end{table}
    \bigskip

    \subsection{Drivers of Non-Axisymmetry}
    \label{sec:drivers_non-axisymmetry}
    
    Many mechanisms  have been proposed to drive non-axisymmetry in dynamo models. 
    The dynamo number $D$ (equation~\ref{D}) or the critical dynamo number $D\crit$
    (equation~\ref{Dcrit}) may vary between the optical spiral arm and inter-arm regions.
    Non-axisymmetric saturated large-scale magnetic field solutions can be obtained 
    by assuming that a parameter in these expressions, 
    such as the turbulent speed $u$ or the vertical outflow speed $U\f$, 
    is spirally modulated.
    Large-scale spiral streaming motions \citep{Moss98} are also possible.  
    Spiral modulation of the equipartition field $B\eq$ 
    can also lead to non-axisymmetric saturated fields \citep{Chamandy+13a} in the nonlinear regime.
    There are other possibilities too (see also Sec.~\ref{sec:phase_shift}).
    The question  is not whether dynamo models can produce non-axisymmetry in saturated large-scale fields,
    but which model best explains the observations?
    
    \subsection{Multiplicity of Magnetic Arms in Dynamo Models}
    \label{sec:multiplicity}
    
    When the dynamo is forced by modulation of $\alpha$ along a steady and rigidly rotating spiral,
    eigen modes with $m=\{n$, $2n$, $3n$, $\ldots\}$ for a galaxy with an $n$-fold spiral structure
    are enslaved to the fastest growing mode, $m=0$, so have the same kinematic growth rate. 
    For realistic parameter values, this set of enslaved non-axisymmetric modes
    grows the fastest, although other sets of enslaved modes can also have significant growth rates,
    e.g. odd modes, enslaved to $m=1$ in a dynamo forced by an $n=2$ spiral \citep{Mestel+Subramanian91,Subramanian+Mestel93}.
    Any asymmetry in the two spiral arms would generate $n=1$ forcing and help to promote $m=1$.
    Studies have explored the non-linear regime, various choices of the modulated parameter, 
    and patterns of spiral modulation that are steady or transient
    \citep{Moss95,Moss98,Rohde+99,Chamandy+13a,Chamandy+13b,Moss+13,Chamandy+15,Moss+15}.
    For realistic parameter values, spiral patterns of multiplicity $n$ 
    lead to significant non-axisymmetric mean magnetic fields with a dominant $m=n$ spiral structure in the saturated regime.
    However, the $m=0$ component typically dominates over components with $m\ne0$ \citep[e.g.][]{Chamandy+14a}.
    
    \subsection{Pitch Angles and Radial Extents of Magnetic Arm Structures in Dynamo Models}
    \label{sec:magnetic_arm_extent}
    
    The pitch angles and radial extents of mean magnetic field spiral arm \textit{structures} 
    in saturated non-axisymmetric mean-field dynamo solutions 
    depend weakly on  which parameter is spirally modulated with  non-axisymmetry, 
    but strongly on the evolution of the forcing spiral, e.g. steady and rigidly rotating vs. transient and winding up.
    If the forcing spiral is rigidly rotating, then magnetic arms cut across forcing arms near the corotation radius,
    with a small pitch angle $\lesssim 10^\circ$ \citep[e.g.][]{Chamandy+13a}.
    This pitch angle is almost independent of the pitch angle of the forcing spiral arms.
    Furthermore, the amplitude of the non-axisymmetric field is sharply peaked at or near corotation.
    Spiral forcing that invokes multiple spiral patterns, 
    each with its own multiplicity and corotation radius,
    can produce more elongated magnetic structures  with  larger overall pitch angles \citep{Chamandy+14a}, 
    but cannot generate the degree of alignment between magnetic and forcing spirals 
    that is inferred from some observations such as those of NGC~6946.
    In contrast, if the forcing spiral is allowed to wind up with time, 
    then the dynamo adjusts rapidly so that the pitch angle of the magnetic arm structure tends to be similar
    to that of the forcing spiral at all times,
    leading to their mutual alignment.
    Magnetic arms then become more extended than in the rigidly rotating case, 
    and resemble those observed \citep{Chamandy+13a,Chamandy+15}.
    This provides a ``magnetic voice'' to a long standing debate, supporting  the idea that spiral arms in galaxies do,
    in fact, wind up, and thus are transient \citep{Dobbs+Baba14,Masters+19}.
    Here again, we see the use of magnetic fields to probe other physics.
    
    \subsection{Localization of Magnetic Arms vis-\`{a}-vis Spiral Arms in Dynamo Models}
    \label{sec:phase_shift}
    
    That magnetic arms are offset from optical arms in azimuthal angle in most cases suggests two possible explanations: 
    (i) the mean-field dynamo is strongest in the spiral arms but the field gets phase-shifted, 
    or (ii) the mean-field dynamo is strongest in the inter-arm regions.
  
    Tentative support for the first idea comes from the analysis of \citet{Frick+00}, who found  that the magnetic arms in NGC~6946 
    are ``phase-shifted images'' of the optical arms, with a negative phase shift with respect to  the rotation.
    Time-delays can cause phase-shifts between different spiral tracers.
    For example, inside the corotaton radius,  the spiral traced by HI surface density should lag that traced by star-formation rate surface density.
    \citet{Chamandy+13a} and \citet{Chamandy+13b} 
    identified a time delay mechanism within dynamo theory to produce steady state azimuthal phase shifts between the magnetic spiral arm pattern 
    and the spiral arm pattern.  The mean-field dynamo requires a finite time, of order the turbulent correlation time $\tau\sim10\Myr$, 
    to respond to changes in the mean magnetic field and small-scale turbulence which  can cause  negative phase shifts of order $\Omega\pat\tau$
    with respect to the forcing spiral, where $\Omega\pat$ is the spiral pattern speed.
    For a steady and rigidly rotating spiral, this implies phase shifts of up to $\sim-30^\circ$ near the corotation radius, 
    where magnetic arms are strongest. 
    However,  this ``$\tau$ effect'' or memory effect has so far not been shown to lead to a phase shift for the case of a transient spiral, 
    which, as noted above, alignment of magnetic and optical arms seems to require.
    
    The second idea, that the dynamo could  be stronger in the optical inter-arm regions,
    naturally leads to  maxima in central field strengths therein, 
    as seen in NGC~6946, but has trouble explaining smaller offsets.  There  exist plausible mechanisms that cause stronger dynamo action in the inter-arm regions \citep{Shukurov98}.
    For example, a larger mean vertical outflow in the spiral arms (due to higher rates of star formation)
     would reduce dynamo action there (Sec.~\ref{sec:strength_theory}).
    If the forcing spiral is modeled as a transient density wave \citep{Binney+Tremaine08},
    this reasonably reproduces the kind of interlaced spiral pattern seen in NGC~6946 \citep{Chamandy+15}.
    Another possibility is a higher rms turbulent velocity in the spiral arms, 
    which would weaken the dynamo through enhanced turbulent diffusion \citep{Moss98,Moss+15}.
    This possibility is explored in Sec.~\ref{sec:veloc_disp_contrast}.
    
    Yet another possibility was explored by \citet{Moss+13}, 
    who injected random mean magnetic field fluctuations into the spiral arms (modeled to be rigidly rotating), 
    and found that the ratio of large-scale to small-scale field strengths -- defined using a Gaussian filter with width $500\pc$ -- turns
    out to be significantly greater in the inter-arm regions than in the arms.
    This  demonstrates that if turbulent tangling/small-scale dynamo action is stronger in the spiral arms, 
    the ratio of regular to turbulent field is larger in the inter-arm regions.
    \citet{Chamandy+Singh18} found in their 1-D model that if the value of $b/B\eq$ at early times, 
    when $\mean{B}/B\eq\ll1$, is larger in the spiral arms due to stronger fluctuation dynamo action there,
    then this leads to both $b/B\eq$ being smaller and $\mean{B}/B\eq$ being larger 
    in the inter-arm regions than in the arms in the steady state.  
    This result has not yet been tested using a global model.
    
    \subsection{Constraints on Non-Axisymmetric Dynamo Models Using Non-Magnetic Galaxy Data}
    \label{sec:veloc_disp_contrast}
    
    Non-magnetic data can be \textit{as essential as magnetic data}
    for constraining dynamo models, by directly constraining input parameters of the model.
    One example is the value of the rms turbulent speed $u$, 
    approximately equal to $\sqrt{3}$ times the 1-D velocity dispersion $\sigma\turb$ for isotropic turbulence.
    Because an azimuthal variation of $u$ has been invoked in models to explain magnetic arms,
    exploring whether a variation of $\sigma\turb$ between arm and inter-arm regions
    exists in galaxies is useful.  We present such a study below, using WISE and THINGS data.
    Our study is demonstrative and preliminary, focussing on a few interesting galaxies,
    but we hope it will encourage more comprehensive interdisciplinary efforts in the future.
    
    \subsubsection{\hi\ Data Products}
    \label{sec:HI_products}
    
    In this work, new \hi\ total intensity and \hi\ velocity dispersion maps were generated for the galaxies M\,51 (NGC\,5194), M\,74 (NGC\,628), and NGC~6946; from the naturally-weighted \hi\ data cubes from The \hi\ Nearby Galaxy Survey (THINGS, \citealt{THINGS_Walter}).  The native properties of these cubes are presented in Table~\ref{cube_properties}. 
    \begin{table}[h]
    \begin{center}
    \caption{Native properties of THINGS \hi\ data cubes}
    \label{cube_properties}
    \begin{tabular}{lccccc}
    \hline
    Galaxy  & $B_\mathrm{maj}$    & $B_\mathrm{min}$ & Noise         & Pixel scale   & Channel width     \\
    		& [$''$]		      &	[$''$]	        & [mJy/beam]    & [$''$]	    & [$\kms$]          \\
    \hline
    M\,51 (NGC\,5194)    &	11.92	&	10.01	&	0.39	&	1.5	&	5.2	\\
    M\,74 (NGC\,628)     &	11.88	&	9.30	&	0.60	&	1.5	&	2.6	\\
    NGC\,6946	&	6.04	&	5.61	& 	0.55	&	1.5	&	2.6	\\
    \hline
    \end{tabular}
    \footnotesize
    \begin{itemize}
    \item[]
    Column 1: galaxy name; columns 2/3: major and minor axis of synthesised beam; column 4: rms noise;\\
    column 5: pixel size; column 6: channel width.
    \end{itemize}
    \end{center}
    \normalsize
    \end{table}
    
    For each \hi\ data cube, the following steps were carried out to produce \hi\ total intensity and \hi\ velocity dispersion maps:\vspace{0.3cm}
    
    \begin{enumerate} 
        \item A Gaussian kernel was used to smooth the cube to a spatial resolution of $13.5''\times 13.5''$.\vspace{0.3cm}
        \item The smoothed cube was spatially re-gridded to have pixels of size $4.5''\times 4.5''$.\vspace{0.3cm}
        \item The mean, $\mu$, and standard deviation, $\sigma$, of the noise in a line-free channel of the smoothed, re-gridded cube was measured.\vspace{0.3cm}
        \item All line profiles in the RA-Dec plane of this cube were fit with a Gaussian.\vspace{0.3cm}
        \item Fitted line profiles with less than 20\% of their flux above a level of $\mu + 2\sigma$ were discarded.\vspace{0.3cm}
        \item The fitted Gaussian parameters of the remaining profiles were used to generate the \hi\ maps.
    \end{enumerate}
    \vspace{0.3cm}
    
\begin{figure*}
    \centering
    \includegraphics[width=1.\columnwidth,angle=0]{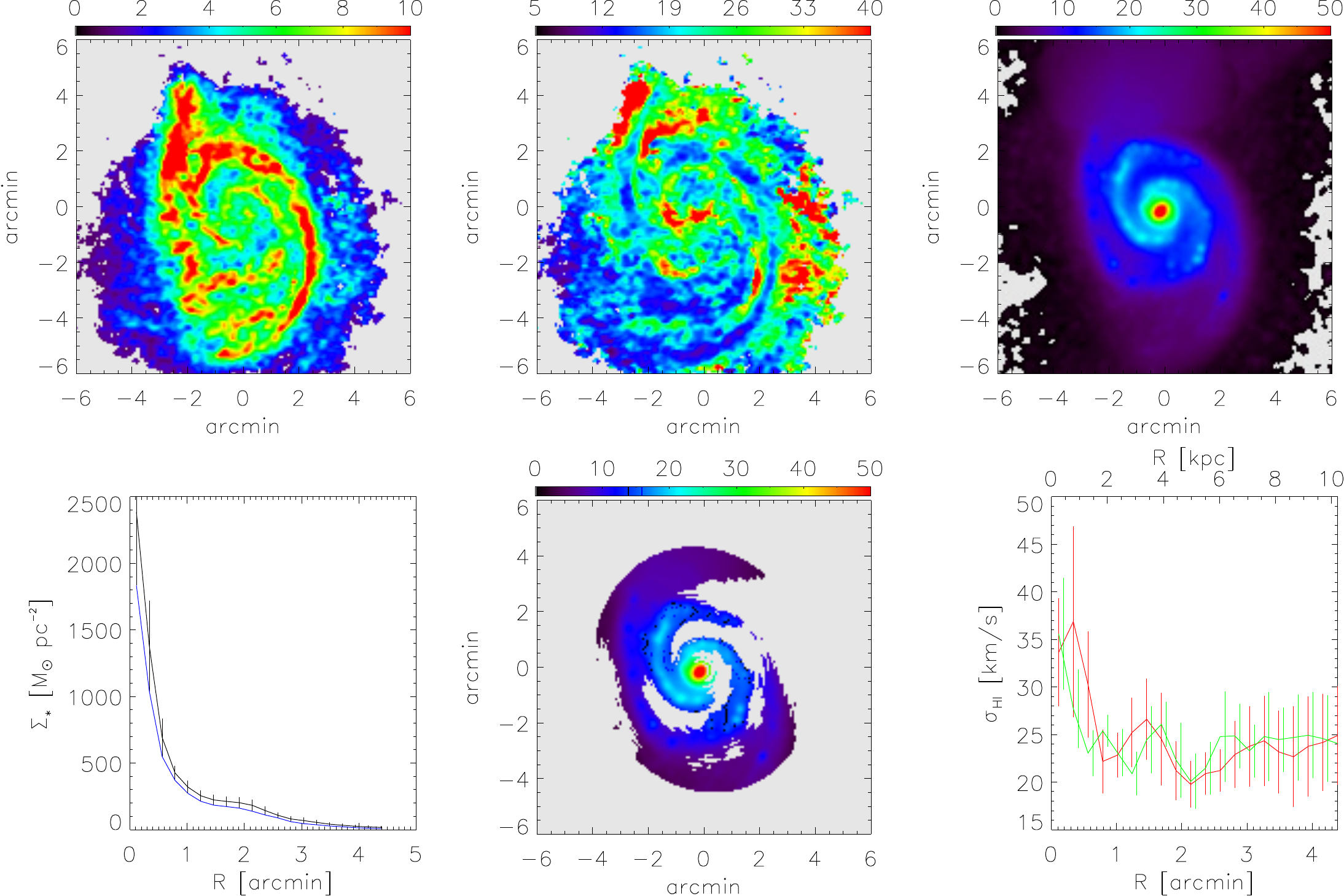}
    \caption{Maps and radial profiles for M\,51 (NGC\,5194).  Top left: \hi\ total intensity map in units of \msun~pc$^{-2}$, top middle: \hi\ velocity dispersion map in units of $\kms$, top right: stellar mass surface density map in units of \msun~pc$^{-2}$, bottom left: azimuthally-averaged stellar mass surface density (black) with the lower limits of the error bars (blue), bottom middle: thresholded stellar surface density map showing arm (colour scale, in units of \msun~pc$^{-2}$) and inter-arm (grey scale) portions of galaxy, bottom right: azimuthally-averaged \hi\ velocity dispersions for arm (red) and inter-arm (green) portions of galaxy. Error bars represent the interquartile range of surface densities in each ring.}
    \label{NGC5194}
    \centering
    \end{figure*}

\begin{figure*}
    \centering
    \includegraphics[width=1.\columnwidth,angle=0]{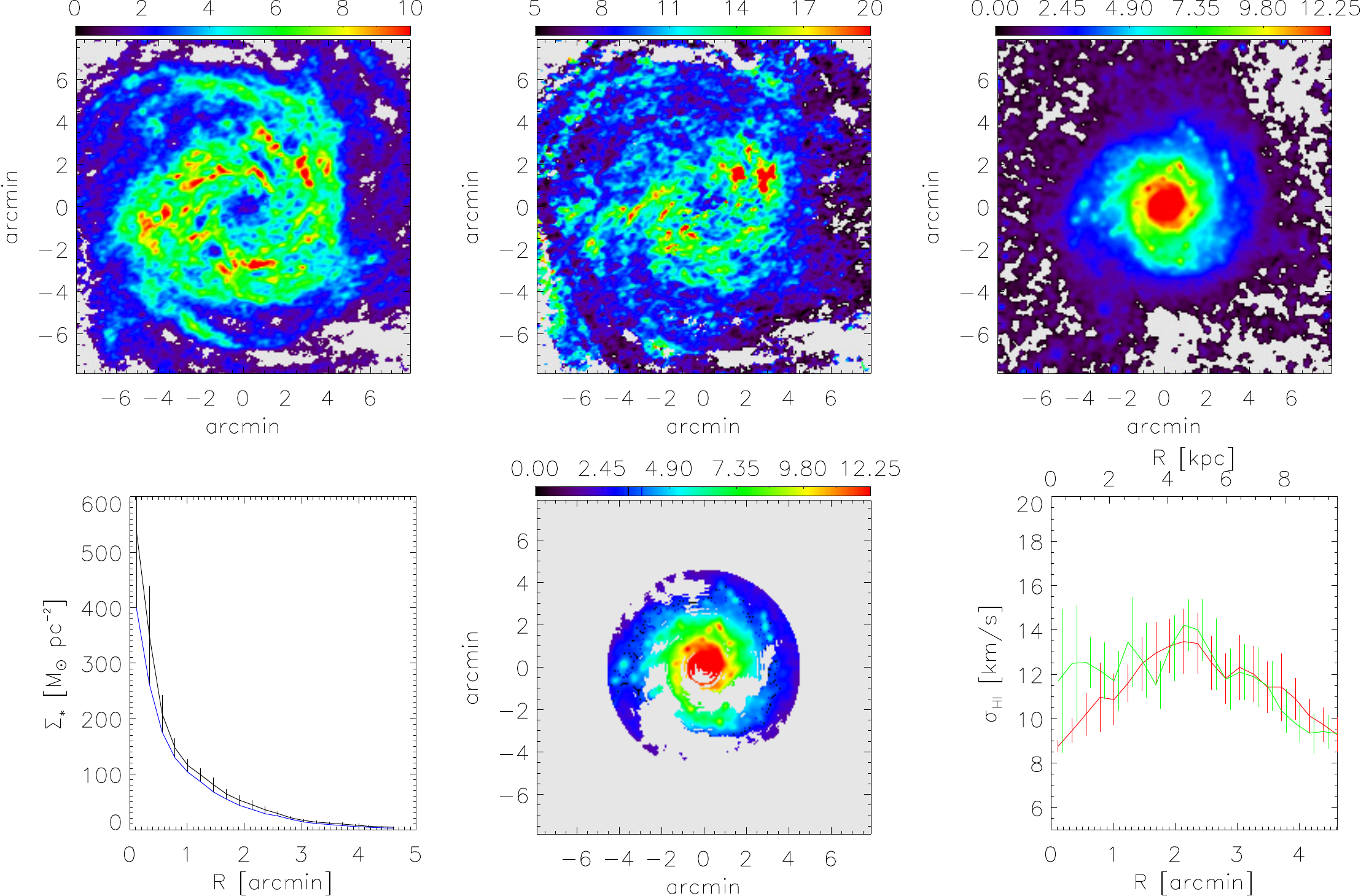}
    \caption{Maps and radial profiles for M\,74 (NGC\,628); see Fig.~\ref{NGC5194} caption for full details.}
    \label{NGC628}
    \centering
    \end{figure*}

\begin{figure*}
    \centering
    \includegraphics[width=1.\columnwidth,angle=0]{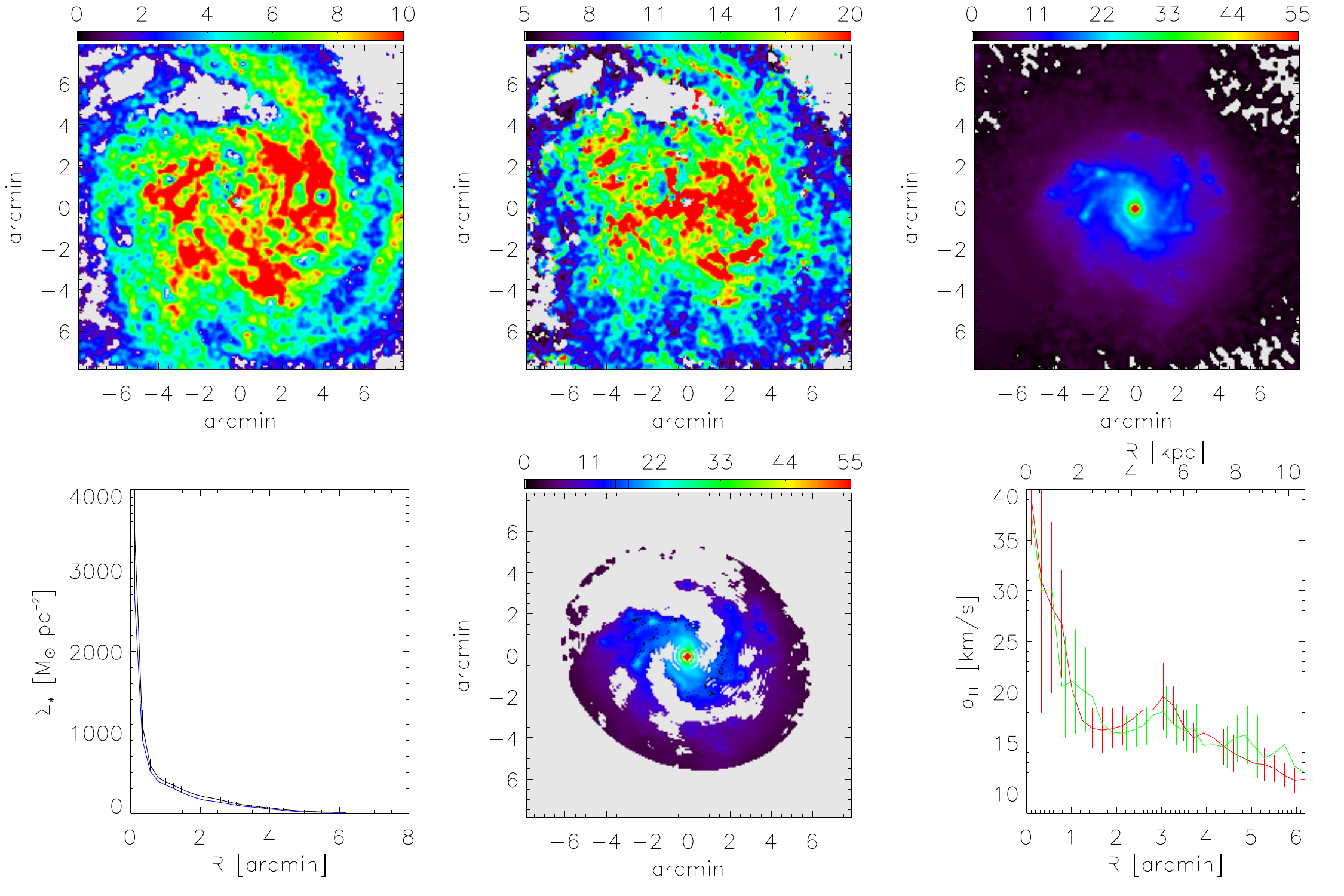}
    \caption{Maps and radial profiles for NGC\,6946; see Fig.~\ref{NGC5194} caption for full details.}
    \label{NGC6946}
    \centering
    \end{figure*}

    The \hi\ total intensity maps and \hi\ velocity dispersion maps for NGC~5194, NGC~628, and NGC~6946 are shown in the top left and top middle panels of Figures \ref{NGC5194}, \ref{NGC628}, and \ref{NGC6946}, respectively.
    
    \begin{figure*}
    \centering
    \includegraphics[width=1.\columnwidth,clip=true,trim= 0 0 0 210,angle=0]{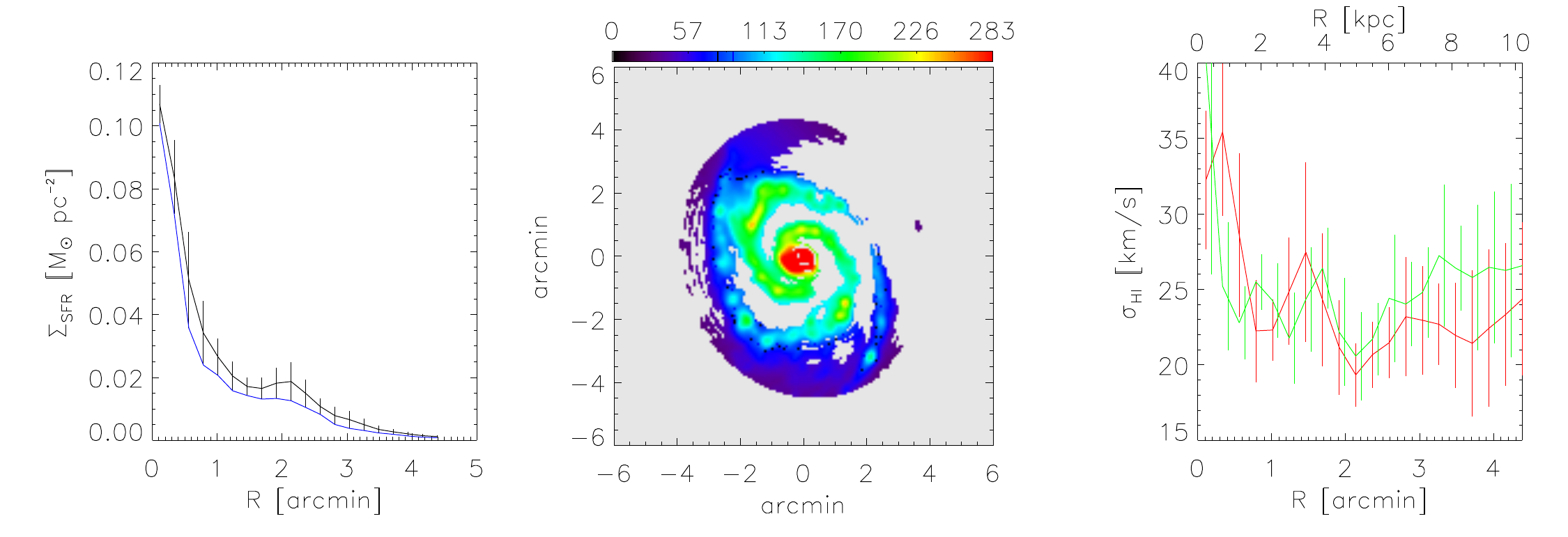}
    \includegraphics[width=1.\columnwidth,clip=true,trim= 0 0 0 210,angle=0]{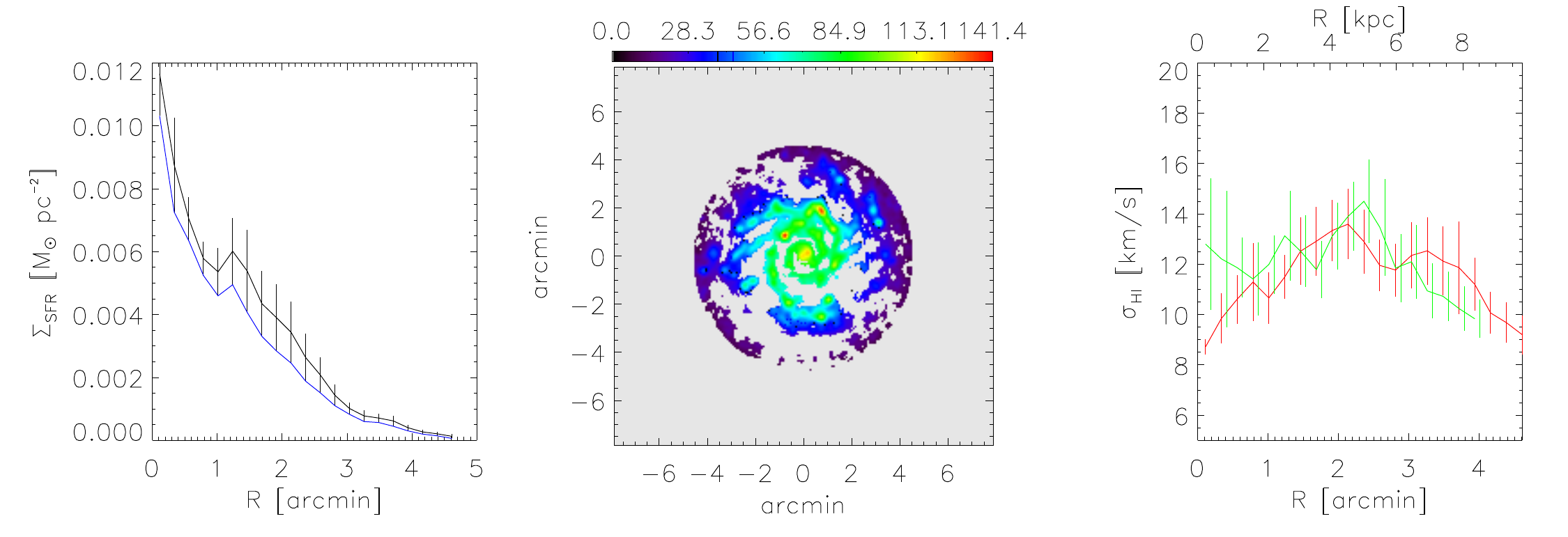}
    \includegraphics[width=1.\columnwidth,clip=true,trim= 0 0 0 210,angle=0]{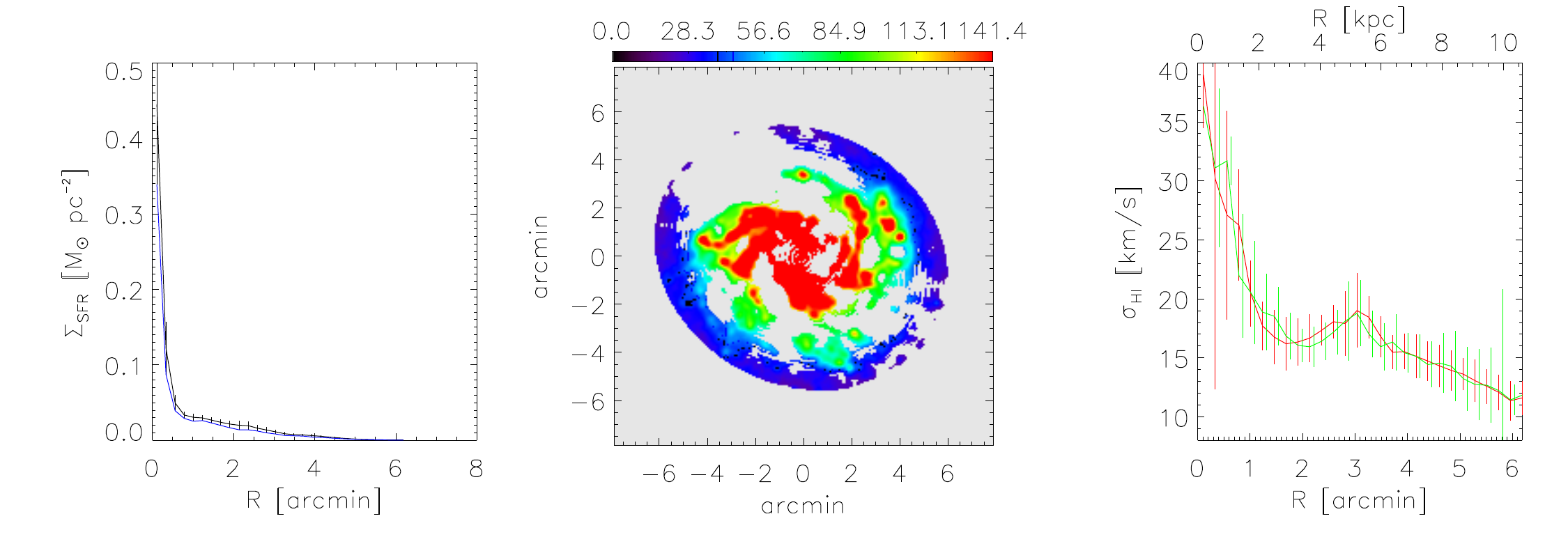}
    \caption{Each row is similar to the bottom row of Figs.~\ref{NGC5194}--\ref{NGC6946},
    but now WISE 12\,$\mu$m data, which traces star formation rate,
    is used to define arm and inter-arm regions.
    Rows from top to bottom show NGC~5194, NGC~628, and NGC~6946.}
    \label{fig:SF}
    \centering
    \end{figure*}

    \subsubsection{Stellar Surface Density Maps}
    
    Stellar mass surface density maps of the galaxies were generated using 3.6\,$\mu$m imaging from the Wide-field Infrared Survey Explorer (WISE, \citealt{WISE_Wright}).  The 3.6\,$\mu$m maps were placed on the same astrometric grid as the \hi\ maps by smoothing them to a spatial resolution of 13.5$''$ and then re-gridding to a pixel size of 4.5$''$.  These modified 3.6\,$\mu$m maps are shown in units of \msun~pc$^{-2}$ in the top right panels of Figures~\ref{NGC5194}, \ref{NGC628}, and \ref{NGC6946}.
    
    For each galaxy, the average disc inclination and position angle from the tilted ring models of \citet{THINGS_deBlok} (given in their Table 2) was used to specify a set of concentric rings of width 13.5$''$.  All rings were centred on the dynamical centre of the galaxy, also measured in \citet{THINGS_deBlok}. The bottom left panels in Figures~\ref{NGC5194}, \ref{NGC628}, and \ref{NGC6946} show as black curves the azimuthally-averaged stellar mass surface densities of the galaxies, in units of \msun~pc$^{-2}$. The blue curves trace the lower limits of the errors bars. 
    For each galaxy, all 3.6\,$\mu$m emission above this level (blue curve) was treated as belonging to the spiral arms of the galaxy, whereas emission below this level was considered to be inter-arm emission. (The results do not change significantly when the upper limits of the error bars are instead used.) The bottom middle panels in Figures~\ref{NGC5194}, \ref{NGC628}, and \ref{NGC6946} show the spiral arm components of each galaxy, as determined using the above-mentioned method. 
    
    \subsubsection{Arm/Inter-Arm \hi\ Velocity Dispersions}
    \label{sec:velocity_dispersion}
    
    A galaxy's 3.6\,$\mu$m emission is a tracer of its old, evolved stellar population constituting the bulk of its stellar mass.  Having decomposed each galaxy's 3.6\,$\mu$m emission into spiral arm and inter-arm regions, we proceeded to measure the typical \hi\ velocity dispersion $\sigma\turb$ in each of the regions.
    Note that for NGC~6946,
    which contains the classic example of interlaced magnetic and optical arms, 
    the optical arms traced by \citet{Frick+00} were in R band, and hence also trace old stars.
    
    The bottom right panels in Figures~\ref{NGC5194}, \ref{NGC628}, and \ref{NGC6946} show the azimuthally-averaged \hi\ velocity dispersions for the arm (red curves and error bars, showing the inter-quartile range) and inter-arm (green curves and error bars) for the galaxies NGC\,5194, NGC\,628, and NGC\,6946, respectively.  While there do exist differences of the order of a few $\kms$ in the azimuthally-averaged \hi\ velocity dispersions for each galaxy, the two profiles are almost always consistent with one another within the uncertainties.
    
    The above-mentioned procedure of decomposing the disc into arm and inter-arm regions was repeated 
    for the galaxies using the WISE 12\,$\mu$m maps, 
    which serve as effective monochromatic tracers of the total star formation rate \citep{Elson_M33_2019}.
    Velocity dispersion might be expected to correlate with star formation rate density \citep[e.g.][]{Schober+16}
    since higher star formation rate density would lead to higher SN rate (SNR) density.
    This case is shown in Fig.~\ref{fig:SF}.
    Each row is similar to the bottom row of Figs.~\ref{NGC5194}--\ref{NGC6946},
    but now using the WISE 12\,$\mu$m to define arm and inter-arm regions.
    The results using this spiral tracer are very similar, 
    which tells us that the results are not sensitive to the spiral tracer used.
    
    Regardless of the map used to identify the spiral structure, 
    the arm and inter-arm values are consistent to within the extents of the error bars.
    This suggests that there is no significant difference between arm/inter-arm \hi\ velocity dispersions. 
    This is also supported by the fact that the difference in the azimuthally averaged $\sigma\turb$
    between arm and inter-arm regions changes sign along radius in each galaxy.
    On the other hand, trends along radius do emerge, 
    but the difference in azimuthally averaged $\sigma\turb$ between arm and inter-arm regions rarely exceeds $20\%$.
    The implication is that azimuthally averaged rms turbulent speed 
    is the same in arm and inter-arm regions to within about $20\%$. 
    
    This is a first attempt at measuring the rms turbulent velocity contrast between arm and inter-arm regions.
    In the future, arm and inter-arm regions could be distinguished by more sophisticated methods, such as wavelet analysis. 
    Furthermore, in order to better constrain the rms turbulent velocity in the disc,
    an attempt could be made to separate out the contribution to the velocity dispersion from thermal broadening \citep{Tamburro+09}.
    
    \subsubsection{Radial variation of the velocity dispersion}
    \label{sec:velocity_dispersion_radial}
    
    Other investigators have measured the radial variation of \hi\ velocity dispersion in THINGS galaxies.  Perhaps most relevant to our efforts are the results of \citet{Mogotsi+16}. In their Fig.~10, they present azimuthally-averaged \hi\ velocity dispersions for 14 THINGS galaxies, including NGC\,628, NGC\,5194, and NGC\,6946.  \citet{Mogotsi+16} measured \hi\ velocity dispersions in two ways: 1) by calculating the second-order moments of the \hi\ line profiles, and 2) by fitting a Gaussian to each line profile and using the standard deviation (as we did in this work).  For both sets of estimates, their radial profiles are similar to ours, and certainly consistent within the errors.  For NGC\,5194 and NGC\,6946, their results (like ours) show the \hi\ velocity dispersion to be highest at the centre of the galaxy, and to decrease with increasing radius.  \citet{Mogotsi+16} indeed find this sort of trend for most of the THINGS galaxies in their sample. 
    Gas velocity dispersions are expected to be highest in the regions of most active star formation given that the energy and momentum output from young, hot stars contributes significantly to the thermal state and bulk motions of the gas.  
    The azimuthal average of the \hi\ velocity dispersion is found to decrease radially much more slowly than the observed star formation rate, traced using infrared and UV emission \cite{Tamburro+09,Ianjamasimanana+15}.  A similar trend is seen here for NGC\,628, NGC\,5194, and NGC\,6946, indicating the dominance of star formation over other sources of turbulence in the inner parts of galaxies.
    
    When the size of an \hi\ beam is large with respect to the angular size of a galaxy with a steeply-rising rotation curve, beam smearing can result in \hi\ velocity dispersions being over-estimated at the centre of a galaxy.  However, for the case of the THINGS data, the beam is small enough relative to the spatial extent of the galaxies to ensure that beam smearing is not contributing to the high \hi\ velocity dispersions we (and \citeauthor{Mogotsi+16}) measure near the centres of NGC\,5194 and NGC\,6946. NGC\,628 has its highest velocity dispersions located at intermediate galactocentric radii, rather than at the centre of the galaxy.  However, this could be due to the prominent \hi\ under-density near the centre of the galaxy.  
    
    \subsubsection{Preliminary theoretical interpretation}
    
    Is variation in $u$ between arm and inter-arm regions expected on theoretical grounds?
    Firstly, we expect $u$ to increase with the SNR but decrease with gas density \citep[e.g.][]{Schober+16}.
    These two effects could offset since spiral arms have both higher SNR and higher gas density.
    
    Secondly, suppose the extreme case, 
    where the SNR is zero between arms and turbulence is driven only in the arms.
    Even then, turbulence would homogenize $u$ across the arms and inter-arms if  turbulent gas driven in the spiral arms can be advected to the adjacent inter-arm region before it decays
    (with decay timescale $\tau\eddy$).
    This condition is 
    \[
      \tau\eddy> \frac{\pi}{n|\Omega-\Omega\pat|} \approx 40\Myr \left(\frac{n}{4}\right)^{-1} \left(\frac{|\Omega-\Omega\pat|}{20\kmskpc}\right)^{-1},
    \]
    where $\tau\eddy=l/u$ is the eddy turnover time (which may in general be different from the correlation time $\tau$), 
    $n$ is the number of equally spaced spiral arms,  and $\Omega\pat$ is the spiral pattern angular speed.
   This inequality is more  likely to hold at radii far from corotation, 
    where $|\Omega-\Omega\pat|$ is large,
   for large $n$, since arms and inter-arms are more closely spaced, and for large $\tau\eddy$.
    
    Thirdly, the turbulence may be driven by the magneto-rotational instability (MRI) 
    or at least MRI might set a lower limit for the value of $u$ \citep{Sellwood+Balbus99}, 
    and MRI-driven turbulence need not be stronger in arms compared to inter-arms.
    
    \subsection{Overall Level of Correspondence with Mean-Field Dynamo Models}
    \label{sec:nonaxisym_summary}
    
    Any of the multiple viable mean field models for the large scale field must explain three broad properties:
    (i) the existence of non-axisymmetric modes (e.g. $m=1$, $m=2$) that approach the strength of 
    the axisymmetric mode ($m=0$);
    (ii) the typical offset of magnetic arms from optical arms, 
    (iii) the several kpc length of magnetic arms and structure pitch angles comparable to those of the optical arms.
    
    Modeling shows that (i) is attainable without fine tuning
    if the mean-field dynamo is \textit{forced} non-axisymmetrically.
    \textit{How} the dynamo is forced is not well understood because, as discussed above, 
    there are multiple options that work to reproduce basic observed features.
    These effects are not mutually exclusive; different effects may 
    be important in different galaxies, and more than one such effect could operate simultaneously.
    To distinguish between the various models, 
    \textit{direct} observation of the  variation of the proposed driver
    (or its observable proxy) between arm and inter-arm regions is desirable.
    This was attempted above for the quantity $u$ (proxy $\sigma\turb$).
    We found such variation to be at most $\sim20\%$ for the three galaxies explored, 
    which would impose important constraints on dynamo models.
    More detailed studies are needed to confirm these preliminary findings.
    The role played by the small-scale magnetic field in large-scale field evolution is largely unexplored, 
    and it may contribute to non-axisymmetric large-scale fields
    when the efficiency of the fluctuation dynamo/turbulent tangling is spirally modulated.

    Property (ii) implies that models must be able to reproduce 
    not only the observed offsets between magnetic and optical arms,
    but the variability in offsets between galaxies or within a galaxy.
    Whether such variability could be produced by variability
    in the underlying spirally-modulated parameter of the simplest non-axisymmetric dynamo models or whether 
    the more complex model ingredients and forcing described above are needed, is not yet clear.
    Nevertheless, large, small, and zero offsets can be reproduced 
    by existing different mean-field dynamo models. 
    To better constrain these models, comparatively
    applying them to one or more specific galaxies, 
    with input parameters constrained by data (along the lines of Sec.~\ref{sec:velocity_dispersion}) would be helpful.
    
    The need to explain property (iii) has  already falsified a certain class of models. 
    Explaining similar structural pitch angles and extents of magnetic and optical arms
    requires the spiral that forces the dynamo to wind up, not rotate rigidly with a constant pattern angular speed.
    Property (iii) seems to be insensitive to other aspects of the models.
    
    \section{Conclusions and Outlook}
    \label{sec:conclusion}
    
    We have reviewed observations and theory of magnetic fields in nearby spiral galaxies,
    with a focus on large-scale fields, namely those  coherent over scales larger than the correlation length of turbulence.
    For the first time, we have consistently computed various important properties of galactic magnetic fields for nearby spiral galaxies that are also relevant for dynamo theory. 
    We have also discussed how these large-scale fields can tell us about other properties of galactic structure.
    
    Magnetic field strength data from spiral galaxies are commonly acquired by assuming that CRs are in local energy equipartition
    with the magnetic field, with a constant ratio of cosmic-ray protons to electrons of $100$. Ref. \citep{Seta+Beck19} provided  support for this assumption on scales above 1\,kpc but not below.  Since using this assumption implies high magnetic field strengths compared to what supersonic MHD turbulence in the absence of CRs would be expected to supply,
    pinning down the accuracy of the assumption is important  for constraining theoretical principles of field origin.    Assuming CR--magnetic field equipartition presumes that a system  has achieved a relaxed minimum energy state, but this may be different from the equilibrium state of a galaxy balanced by a combination of forcing and relaxation.
    
    Mean-field dynamo models can account for the broad global features of such regular fields,
    including strengths of $\sim 1\mkG$, pitch angles of $\sim 5^\circ - 40^\circ$,
    predominance of the $m=0$ azimuthal mode, and the presence of higher modes.
    Other features, such as the locations of magnetic arms vis-\`{a}-vis optical arms,
    can be reproduced by models using reasonable  assumptions, 
    such as a winding up spiral density wave with outflows stronger in the optical arms than between them. 
    However,
    pinning down which  assumptions are most justifiable requires further improvements 
    in observations and theory, and in their mutual comparison. There is a need for simulators, theorists, and observers to give more careful consideration
    to their specific methods of averaging to achieve accurate mutual comparison. 
    This involves developing and applying rigorous tools,
    such as the azimuthal mode analysis to compute the regular field from observations as discussed above,
    and also applying the \textit{same} averaging procedures 
    to theoretical models or simulations.
    Otherwise the ``mean field'' reported by theorists cannot be reliably compared with that reported by observers.
    \footnote{\citet{Zhou+18} 
    explored the general case of comparing models and observations 
    which employ different methods of averaging and highlighted the precision error caused when they are different.
    By using, as far as possible, the \textit{same} averaging method in models as used in observations, 
    their mutual comparison would become more straightforward.}
    
    A basic problem  is that mean field theorists typically assume an infinite ensemble average, 
    equivalent to a volume or time average only for infinite scale separation between fluctuations and mean variation scales, 
    whereas observations probe some version of spatial averages in galaxies where scale separation is finite.
    Addressing this mismatch ultimately requires a model for not just the infinite ensemble averaged magnetic field, 
    but also the fluctuating magnetic field (the total field is the sum of these contributions).
    In mean-field models, a form of averaging that mimics the inherent averaging of observations
    could be incorporated into the mean-field equations
    from the outset (Sec.~\ref{sec:theory_def}) to obtain more \textit{accurate} predictions
    as well as to estimate the theoretical uncertainty (precision) of predictions.
    Alternatively, accounting for the effects of averaging a posteriori 
    may  be possible
    by adding magnetic noise to represent the fluctuating component, 
    and then appropriately averaging to mimic observational averaging.
    As with DNS, the result would still represent  one  statistical realization.
    More accurate predictions and estimates of theoretical uncertainties could be computed 
    by accounting for  distributions obtained over many realizations.
    Other sources of noise besides turbulence,
    such as the locations of SNe in DNS, can also add to the theoretical uncertainty.
   
    \textit{Global} MHD DNSs of galaxies are increasingly common, 
    but the resolution is presently insufficient to capture all of the relevant physics,
    particularly the SN-driven turbulence sourcing large-scale and small-scale dynamo action.
    \textit{Local} ISM simulations in a shearing box can include more physics at small scales
    but neglect global transport.
    Mean-field simulations still have an important role to play, as a bridge between theory and DNS,
    and for explaining or predicting observed global properties.
    At the same time, basic theory and idealized MHD turbulence simulations 
    remain instrumental for studying the underlying physics.
    
    Two areas stand out as particularly promising for synthesizing dynamo theory and observations in the near future.
    First, it is becoming more feasible to model specific galaxies like M~31 or M~51,
    as kinematic and other data used as input into models, improves.
    Such galaxies could be used as testbeds for theory.
    Second, 2-D global axisymmetric, 
    1-D global axisymmetric `no-$z$', or even 0-D local in $r$ axisymmetric `no-$z$' 
    mean-field models can be useful in predicting and explaining 
    observed statistical relations between  various properties of galaxies.
    For example, recent work by \citet{Rodrigues+19a} uses, as input,  semi-analytic galaxy formation models 
    (which themselves use merger-trees from cosmological N-body simulations as input)
    to solve for the magnetic fields of  $\sim 3\times10^6$ galaxies through cosmic time,
    using a 1-D global mean-field dynamo simulation for each evolving galaxy.
    Such ``population synthesis'' models can be valuable tools for 
    generating synthetic statistical data sets.
    
    Specific strategies for improving dynamo models and facilitating their
    synthesis with observational studies are:
    \vspace{0.3cm}
    \begin{itemize}
      \item making averaging in models consistent with averaging in observations or simulations to enable direct comparison;
      \vspace{0.3cm}
      \item modeling parameters of interstellar turbulence as functions of observables 
      using analytical theory and turbulent ISM/galaxy simulations \citep{Hollins+17}, 
      to better constrain dynamo models;
      \vspace{0.3cm}
      \item including better models of small-scale magnetic field from the fluctuation dynamo, turbulent tangling, 
      and helicity conservation, to explain observed isotropic and anisotropic turbulent fields;
      \vspace{0.3cm}
      \item accounting self-consistently for all the effects of this small-scale magnetic field  on the mean electromotive force to obtain more realistic dynamo solutions;
      \vspace{0.3cm}
      \item extending mean-field models to include magnetic feedback onto the mean velocity field
      to better understand phenomena like magnetized outflows and spiral arms;
      \vspace{0.3cm}
      \item quantifying the dependence of the dynamo on the ionization fraction 
      by including partial ionization and ambipolar diffusion to enable more direct comparison with observation;
      \vspace{0.3cm}
      \item including cosmic rays in dynamo models both for their possible role in magnetic field evolution, 
      and to better constrain their properties to enable improved observational estimates of magnetic field properties;
      \vspace{0.3cm}
      \item using global galaxy and local ISM DNS as a complementary laboratory both for testing the theory and synthesizing observations.
      \vspace{0.3cm}
    \end{itemize}
    
    With new radio telescopes like the Square Kilometre Array (SKA) on the horizon,
    large datasets of magnetic and non-magnetic properties from thousands of galaxies with higher angular resolution and sensitivity, 
    many at high redshift and spatially unresolved, will become available.
    Additional measurements in other spectral ranges will constrain important parameters of dynamo theory.
    Some of the most important tasks are:
    \vspace{0.3cm}
    \begin{itemize}
      \item  resolving the structure of tangled/twisted/bent fields and the mysterious anisotropic turbulent fields;
      \vspace{0.3cm}
      \item distinguishing regular from anisotropic turbulent fields with the help of high-quality $RM$ data;
      \vspace{0.3cm}
      \item measuring thermal gas densities from extinction-corrected  H$\alpha$ emission data (or other emission lines), to compute the strength of regular fields from $RM$ data;
      \vspace{0.3cm}
      \item identifying high-order azimuthal modes of regular fields in galaxy discs;
      \vspace{0.3cm}
      \item searching for field reversals in galaxy discs and halos;
      \vspace{0.3cm}
      \item measuring field parities in galaxy discs and halos;
      \vspace{0.3cm}
      \item measuring velocity dispersions in arm and inter-arm regions from \hi\ emission data;
      \vspace{0.3cm}
      \item measuring scale heights of ionized gas discs from H$\alpha$ emission data;
      \vspace{0.3cm}
      \item studying the evolution of large-scale regular fields in galaxies at various redshifts.
      \vspace{0.3cm}
      
    \end{itemize}
    
    Overall, we have realized that progress will require observers and theorists to better ensure 
    that quantities which observers measure and theorists predict are the same. 
    We have pinpointed some of the gaps that must be bridged for this to improve.
    We have also identified new avenues for making direct connections between theory and observations.
    Together, these provide tractable opportunity for progress in understanding galactic magnetic fields.
    
    \bigskip
    
    \abbreviations{The following abbreviations are used in this manuscript:\\
    
    \noindent
    \begin{tabular}{@{}ll}
    MHD & magnetohydrodynamics\\
    DNS & direct numerical simulation\\
    ISM & interstellar medium\\
    SN & supernova\\
    CR & cosmic ray\\
    SFR & star-formation rate\\
    RM & Faraday rotation measures\\
    \end{tabular}}
    
    \bigskip
    
    \authorcontributions{Conceptualization: R.B. and L.C.; radio polarization data and analysis: R.B.; \hi\ data and analysis: E.E.; theory: L.C. and E.G.B.}
    
    \funding{L.C. acknowledges Newcastle University for support during his visit from June 9-30, 2019. L.C. and E.G.B. acknowledge support from NSF Grant AST-1813298.  E.E. is supported by the South African Radio Astronomy Observatory, which is a facility of the National Research Foundation, an agency of the Department of Science and Technology. E.G.B. acknowledges support from KITP UC Santa Barbara, funded by NSF Grant PHY-1748958, and Aspen Center for Physics, funded by NSF Grant PHY-1607611.}
    
    \acknowledgments{We are grateful to Marita Krause, Sui Ann Mao, and the anonymous referees for many useful comments on the manuscript.}
    
    \conflictsofinterest{The authors declare no conflict of interest.}
    
    \newpage
    
    \appendixtitles{yes}
    
    \appendix
    
    \section{Rotation measures in M\,83}
    \label{sec:M83}
    
    M\,83 is a nearby spiral galaxy (at an estimated distance of 8.9\,Mpc) hosting a large stellar bar. Radio polarization data observed with the Effelsberg 100-m telescope at 2.8\,cm wavelength at $75''$ beam width \cite{Neininger+91} and 6.3\,cm wavelength at $2.45'$ beam width \cite{Neininger+93} were combined to construct a map of Faraday rotation measures ($RM$) at $2.45'$ resolution. $RM$ was found to vary double-periodically with azimuthal angle $\phi$ in the galaxy plane, indicative of a large-scale bisymmetric spiral (mode $m=1$) magnetic field \cite{Neininger+93}.

 \begin{figure*}[t]
    \vspace*{7mm}
    \begin{center}
    \hspace{0.2cm}
    \begin{subfigure}{0.5\textwidth}
        \includegraphics[width=7.9cm]{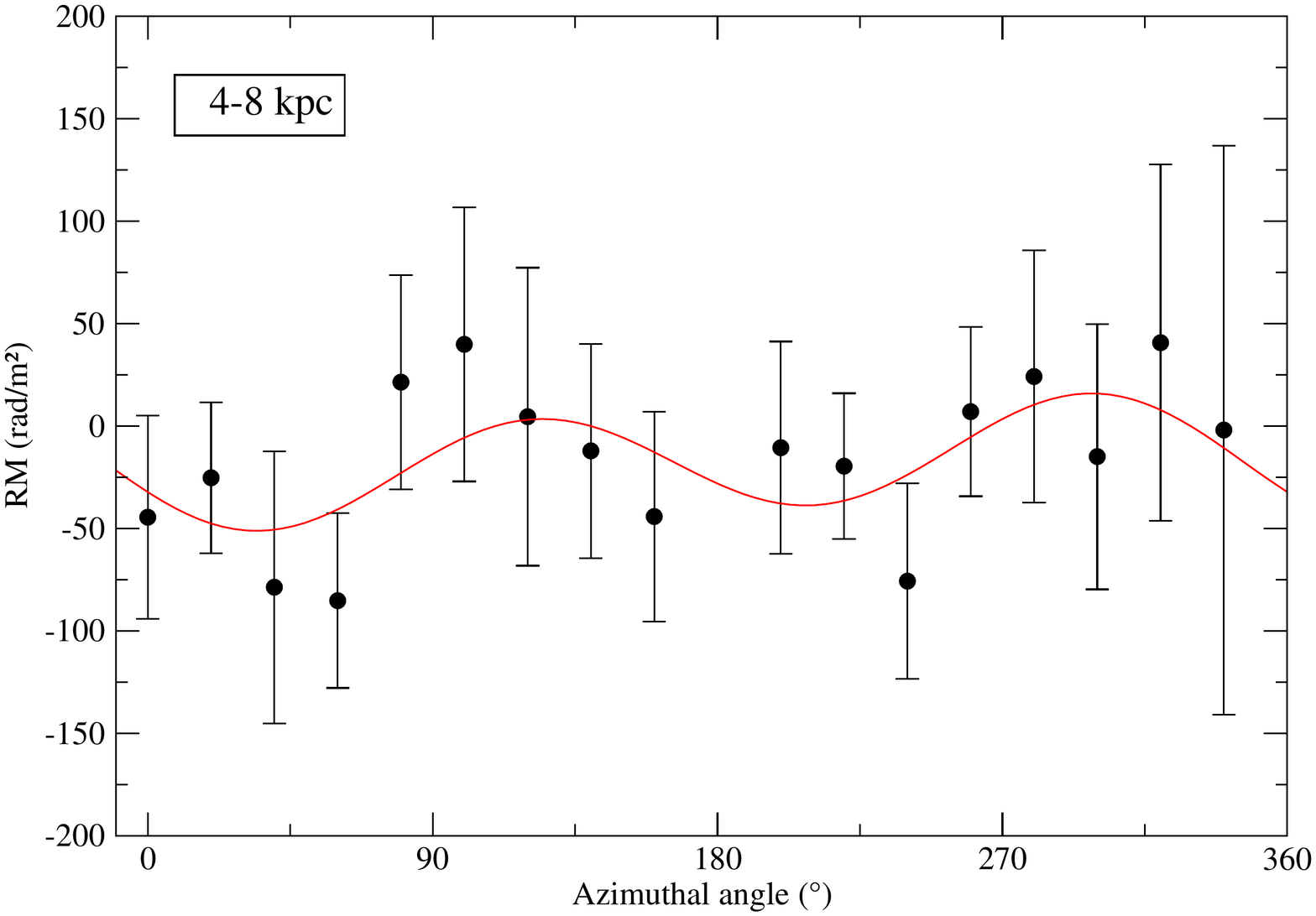}
    \end{subfigure}
    \hspace{-0.5cm}
    \begin{subfigure}{0.5\textwidth}
        \includegraphics[width=7.9cm]{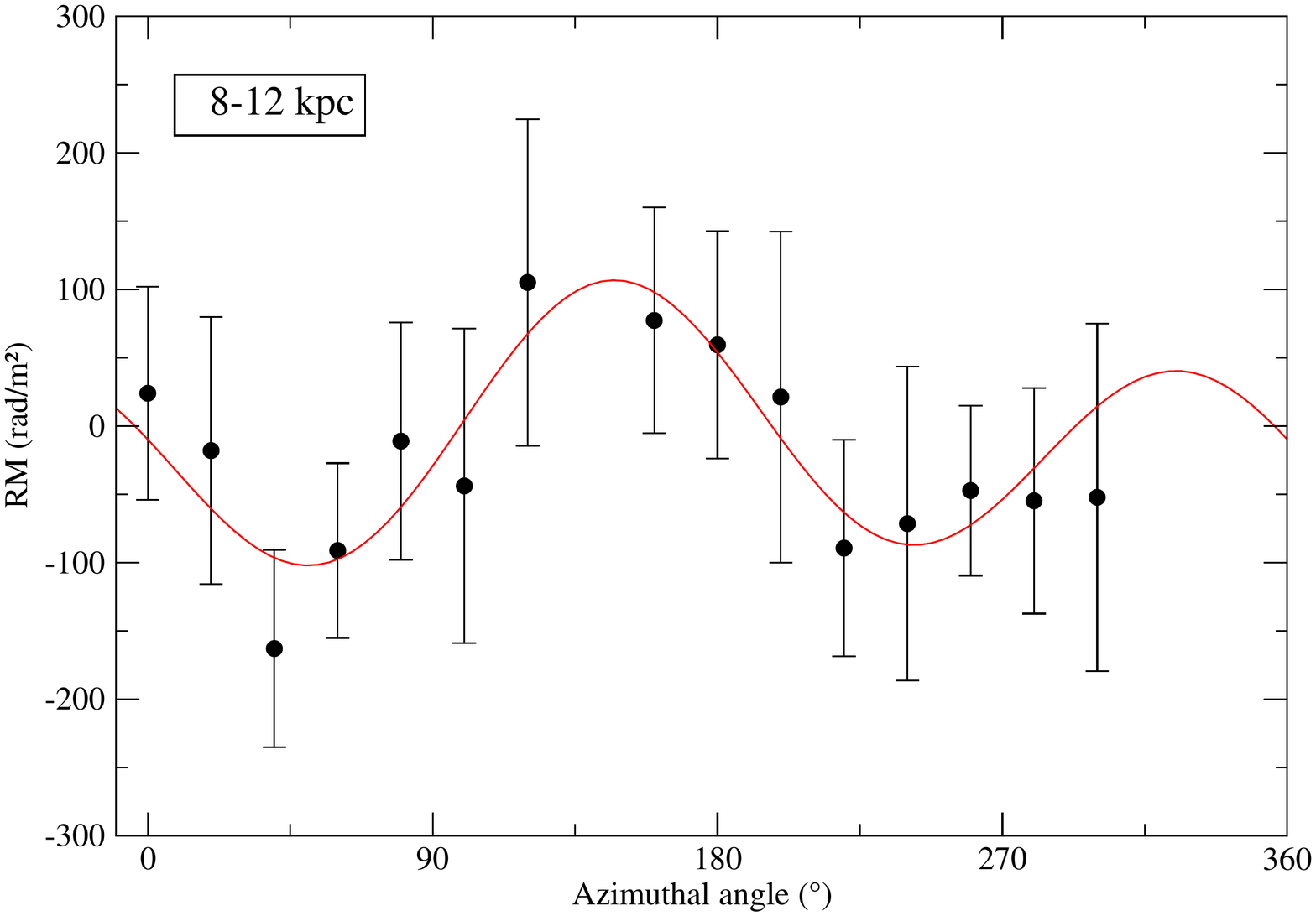}
    \end{subfigure}
    \caption{
    Variation with azimuthal angle of Faraday rotation ($RM$) of the barred spiral galaxy M\,83, measured between 2.8\,cm wavelength \cite{Neininger+91} and 6.3\,cm wavelength \cite{Frick+16}, both at $75''$ beam width, in sectors of two radial ranges 4--8\,kpc and 8--12\,kpc in the galaxy plane (inclined by $24^{\circ}$).
    The azimuthal angle $\phi$ is counted counter-clockwise from the north-eastern major axis (position angle $45^{\circ}$).
    The error bars show the $RM$ dispersion in each sector, including systematic variations, and hence are upper limits of the statistical uncertainties.
    }
    \label{fig:m83}
    \end{center}
    \end{figure*}

     Radio polarization data at 6.1\,cm wavelength observed with the Very Large Array (VLA) with $10''$ beam width were combined with the Effelsberg data at 6.3\,cm wavelength \cite{Frick+16}. This allows us to compute an improved $RM$ map at $75''$ resolution and the azimuthal $RM$ variation at different radii. Significant large-scale variations are found in two radial ranges (Figure~\ref{fig:m83}). The red curves show the fits by a sinusoidal variation (axisymmetric mode $m=0$) plus a double-sinusoidal variation (bisymmetric mode $m=1$), using the formula
    $RM = RM_\mathrm{fg} + RM_\mathrm{0} \, \cos(\phi - p_\mathrm{0}) + RM_\mathrm{1} \, \cos(2 (\phi - p_\mathrm{1}))$, where
    $RM_\mathrm{fg}$ is the foreground $RM$, $RM_\mathrm{0}$ and $p_\mathrm{0}$ are the amplitude and phase of the $m=0$ mode, and $RM_\mathrm{1}$ and $p_\mathrm{1}$ are the amplitude and phase of the $m=1$ mode.
    The parameters and reduced $\chi^2$ values of the fits are given in Table~\ref{tab:m83}.
    Including the mode $m=2$ has not been attempted because of the limited angular resolution.

    The phase $p_\mathrm{1}$ of the $m=1$ mode changes by $(25^\circ\pm15^\circ)+180^\circ=205^\circ\pm15^\circ$ between the two radial ranges. This requires an average pitch angle of the spiral pattern of the $m=1$ regular field of $8^\circ\pm1^\circ$ (using Eq.~A10 in \citet{Krause+89b}), which is consistent with the average pitch angle of the spiral arms of gas and ordered magnetic field as measured by \citet{Frick+16}.
    
    As the signs of $RM$ and radial component of rotational velocity are the same on the major axis (Fig.~\ref{fig:m83}), as previously indicated in the maps by \citet{Heald+16}, the radial component of the axisymmetric field points outward (see Fig.~1 in \citet{Krause+Beck98}).

    \begin{table*}
    \begin{center}
    \caption{
    Fits to the azimuthal variations of Figure~\ref{fig:m83}.
    The phase $p_\mathrm{0}$ corresponds to the average pitch angle of the spiral magnetic field; its small values are consistent with the observed pitch angle of the spiral arms \cite{Frick+16}.
    The reduced $\chi^2$ values smaller than 1 indicate that the error bars are overestimated.
    }
    \label{tab:m83}
    \vspace{0.2cm}
    \begin{tabular}{@{}lcccccc@{}}
    \hline
    Radial range & $RM_\mathrm{fg}$ & $RM_\mathrm{0}$ & $p_\mathrm{0}$ & $RM_\mathrm{1}$ & $p_\mathrm{1}$ & Reduced $\chi^2$ \\
                 &[$\mathrm{kpc}$]  & [rad/m$^2$]     &  [$^{\circ}$]  & [rad/m$^2$]     & [$^{\circ}$]   \\
    \hline
    4--8  & $-18\pm9$  & $-9\pm12$  & $77\pm88$ & $27\pm13$ & $-59\pm14$ & 0.46 \\
    8--12 & $-10\pm13$ & $-34\pm19$ & $-21\pm25$ & $83\pm19$ & $-34\pm6$ & 0.26 \\
    \hline
    \end{tabular}
    \end{center}
    \end{table*}   
    
    \bigskip
    \bigskip

    \reftitle{References}
    
    \bibliographystyle{mnras}
    \bibliography{refs}
    \end{document}